\documentstyle[psfig]{laa}     
\begin{document}
\font\cbf=cmbsy10
\font\cal=cmsy10

%
%
%
\newcommand{\DXDYCZ}[3]{\left( \frac{ \partial #1 }{ \partial #2 }\right)_{#3}}
\newcommand{\pg}{P_{\rm G}}
\newcommand{\pgn}{P_{{\rm G}0}}
\newcommand{\pcr}{P_{\rm C}}
\newcommand{\pcrn}{P_{{\rm C}0}}
\newcommand{\pmag}{P_{\rm B}}
\newcommand{\pmagn}{P_{{\rm B}0}}
\newcommand{\pw}{P_{\rm W}}
\newcommand{\pwn}{P_{{\rm W}0}}
\newcommand{\cg}{c_{\rm G}}
\newcommand{\cgn}{c_{{\rm G}0}}
\newcommand{\fc}{\vec{F}_{\rm C}}
\newcommand{\fcn}{F_{{\rm C}0}}
\newcommand{\fw}{\vec{F}_{\rm W}}
\newcommand{\fwn}{F_{{\rm W}0}}
\newcommand{\ptot}{P_{\rm tot}}
\newcommand{\gag}{\gamma_{\rm G}}
\newcommand{\ggn}{\gamma_{{\rm G}0}}
\newcommand{\gc}{\gamma_{\rm C}}
\newcommand{\gcn}{\gamma_{{\rm C}0}}
\newcommand{\gaw}{\gamma_{\rm W}}
\newcommand{\gwn}{\gamma_{{\rm W}0}}
\newcommand{\va}{v_{\rm A}}
\newcommand{\van}{v_{{\rm A}0}}
\newcommand{\ma}{M_{\rm A}}
\newcommand{\man}{M_{{\rm A}0}}
\newcommand{\mgw}{\dot M_{\rm GW}}
\newcommand{\et}{{\it et al.\/}\,}
\newcommand{\ha}{${\rm H}\alpha$}
\newcommand{\beq}{\begin{equation}}
\newcommand{\eeq}{\end{equation}}
\newcommand{\beqarn}{\begin{eqnarray}}
\newcommand{\eeqarn}{\end{eqnarray}}
\newcommand{\beqaru}{\begin{eqnarray*}}
\newcommand{\eeqaru}{\end{eqnarray*}}
\newcommand{\bfigo}{\begin{figure}}
\newcommand{\efigo}{\end{figure}}
\newcommand{\bpic}{\begin{picture}}
\newcommand{\epic}{\end{picture}}
\newcommand{\bfigd}{\begin{figure*}}
\newcommand{\efigd}{\end{figure*}}
\newcommand{\btabo}{\begin{table}}
\newcommand{\etabo}{\end{table}}
\newcommand{\btabd}{\begin{table*}}
\newcommand{\etabd}{\end{table*}}
%
%
%
%
%
\newcommand{\mG}{\,{\rm mG}}
\newcommand{\muG}{\,\mu\/{\rm G}}
\newcommand{\dyn}{\,{\rm dyn}}
\newcommand{\erg}{\,{\rm erg}}
\newcommand{\parsec}{\,{\rm pc}}
\newcommand{\kpc}{\,{\rm kpc}}
\newcommand{\yr}{\,{\rm yr}}
\newcommand{\second}{\,{\rm sec}}
\newcommand{\msol}{\,{\rm M_{\odot}}}
\newcommand{\msolyr}{\,{\rm M_{\odot}} \, {\rm yr}^{-1}}

   \thesaurus{02.13.2; 
              09.03.2; 
	      09.07.1; 
              09.10.1; 
	      11.08.1; 
              13.25.4  
	      }
   \title{The dynamical signature of the ISM in soft X-rays}

   \subtitle{I. Diffuse soft X-rays from galaxies}

   \author{Dieter Breitschwerdt\inst{1,}\inst{2} \and Thomas
   Schmutzler\inst{3,}$^\dagger$}

   \offprints{D.\ Breitschwerdt} 

   \institute{Max-Planck-Institut f\"ur Extraterrestrische Physik, 
              Postfach 1603, 
              D--85740 Garching, Germany\\ 
              email: breitsch@mpe.mpg.de
	 \and Heisenberg Fellow
         \and Am Beisenkamp 12, D--44866 Bochum, Germany, 
              email: Th.Schmutzler@t-online.de\\
	      $^\dagger$deceased}

   \date{Received 3 August 1998 / Accepted January 25 1999}

   \maketitle

   \begin{abstract}
We present the first dynamically and thermally self-consistent calculations of 
fast adiabatically expanding gas flows from the Galactic disk into the halo. 
It is shown that in a hot plasma ($T \geq 10^6 \,{\rm K}$) with a high overpressure
with respect to the ambient medium, the dynamical time scale is much shorter than 
the intrinsic time scales (e.g.\ for recombination, collisional excitation and 
ionization etc.). Therefore dynamical models that use collisional ionization 
equilibrium (CIE) cooling functions for the evolution of the plasma are 
in general not correct.
In particular, the emission spectra obtained from non-equilibrium calculations are 
radically different. We describe a method to obtain self-consistent solutions 
using an iterative procedure. It is demonstrated that soft X-ray
background emission between 0.3 and 1.5 keV can be well explained by a superposition 
of line emission and delayed recombination of an initially hot plasma streaming 
away from the Galactic disk (outflow and/or winds). 
In addition to these local winds we also present calculations on global winds
from spiral galaxies, which originate from a hot and quiescent galactic corona. 
We also emphasize that it is dangerous 
to derive plasma temperatures merely from line ratios of ionized species, such 
as N\,{\sc v}/O\,{\sc vi}, unless the dynamical and thermal history of the
plasma is known. 
   \end{abstract}
      \keywords{Magnetohydrodynamics (MHD) -- (ISM:) cosmic rays -- 
                ISM: general -- ISM: jets and outflows -- Galaxies: halos 
                -- X-rays: ISM  }

\section{Introduction}
The idea of a hot corona, surrounding our Galaxy and thereby confining
infalling H\,{\sc i} clouds was postulated more than 40 years ago by 
Spitzer (1956). With the discovery of the ubiquitous interstellar 
O\,{\sc vi} absorption line (Jenkins \& Meloy 1974; York 1974) 
by the {\sc Copernicus} satellite, it became plausible that most of 
the interstellar medium (ISM) was filled by a hot and tenuous gaseous 
component, the so-called hot intercloud medium (HIM) consisting mainly of 
interconnected bubbles. Subsequently, 
McKee and Ostriker (1977) calculated the global mass and energy
balance of the ISM, assuming pressure equilibrium between the various
phases including the HIM, which was thought to be maintained by the
energy input of supernova remnants (SNRs). In their global equilibrium model
the hot gas (typical number density
$n \sim 6 \times 10^{-3} \, {\rm cm}^{-3}$ and temperature 
$T \sim 5 \times 10^5 \, {\rm K}$) was not confined to the molecular gas disk 
with a scale height of $\sim 120 \, {\rm pc}$, but would rise into the halo with 
a typical scale height of $H \sim  5 \, (T/10^6 \, [{\rm K}]) \, {\rm kpc}$.

Moreover, part of the HIM (of order of $1 \, \msolyr$) was even thought to
be leaving the gravitational potential well of the Galaxy in the form
of a Galactic wind (see also Mathews \& Baker 1971). However, detailed
numerical simulations by Habe \& Ikeuchi (1980) showed, that a minimum
temperature of $T \approx 4 \times 10^6 \, {\rm K}$ was required
for a thermal Galactic wind, even when the support of centrifugal forces was  
included. Therefore, the bulk of the hot gas that is flowing out of the
disk into the so-called disk-halo connection or the lower halo, would
eventually become thermally unstable due to radiative losses and fall
back to the disk, as it was discussed in the ``Galactic fountain model''
(Shapiro \& Field 1976; Bregman 1980; Kahn 1981). For an alternative view, 
according to which the hot medium remains confined in individual bubbles due 
to an assumed high (magnetic) pressure of the ambient medium, and much 
of the energy is dissipated before break-out can occur, see e.g.\ Cox (1991) 
and Slavin \& Cox (1993). However, as we shall see below, observations 
show the existence of an extended Galactic corona, emitting in soft X-rays; 
it is highly improbable that extragalactic infall alone can account for this, 
and therefore we think that the bulk of hot gas has to be supplied by the disk.

A later version of
the fountain model with a somewhat different flavour is the ``chimney model''
(Ikeuchi 1988; Norman \& Ikeuchi 1989), in which the circulation of 
the gas flow in the lower halo is achieved by channelling the gas 
through pipes (``chimneys''), which are physically connected to underlying
OB-associations. In essence this model represents just a
``clustered Galactic fountain''. Again, the total mass loss rate of
the Galaxy leads to a loss of only a small fraction of the circulated material,
depending on the star formation activity in the Galactic disk. 
The existence of a Galactic disk-halo outflow in the Perseus arm has been confirmed 
by recent observations of Normandeau et al. (1996).
  
It has been pointed out by Johnson \& Axford (1971) and later 
shown by Ipavich (1975) that in principal the dynamic pressure of the
high energy component of the ISM, i.e.\ the cosmic rays (CRs), could
assist in driving a galactic wind, provided that a coupling mechanism
between CRs and gas exists. It can be shown that the escape of CRs from
the Galaxy is accompanied by the resonant generation of small scale MHD
waves via the streaming instability (Lerche 1967; Kulsrud \& Pearce 1969).
Scattering off these
waves reduces the particle drift speed to essentially the Alfv\'en speed
and hence leads to a net forward momentum transfer to the gas via the
frozen-in waves as a mediator. Models with realistic gravitational
potential and geometry (flux tube formalism) for the Galaxy, demonstrate
that CR driven winds with a total mass loss rate of order of $1 \, \msolyr$
are likely to occur in the Milky Way (Breit\-schwerdt et al.\ 1987, 1991). 
In this paper we formally distinguish between \textit{global} winds, that 
originate from the large-scale expansion of a hot galactic corona, 
and \textit{local} winds, which are linked to individual superbubble regions 
in the disk; in both cases the CRs assist in driving the outflow. It is 
clear that in reality also intermediate cases will occur.

Due to their SNR origin, outflows and winds should be best detected in the 
soft Xray regime ($\sim 0.1 - 2.0$ keV). 
Starting with the pioneering observations of Bowyer et al.\ (1968), the
existence of a diffuse soft X-ray component in the Galaxy has been
established. Initially however, it was not clear whether this emission
should be attributed to a purely Local Hot Bubble with an average extension 
of 100 pc (e.g., Tanaka \& Bleeker 1977; Sanders et al.\ 1977), in which the solar 
is system is embedded,  or whether
an extended, diffuse halo component would be present as well. The latter was 
confirmed by the 
Wisconsin All Sky Survey (McCammon \& Sanders 1990, and references therein)
and with much higher sensitivity and spatial resolution by the {\sc Rosat} All 
Sky Survey (e.g.\ Freyberg \& Schmitt 1998; Wang 1998). 
The diffuse nature of the emission was demonstrated by so-called shadowing 
experiments
(Snowden et al.\ 1991; Kerp et al.\ 1993), which were one of the first
deep pointed {\sc Rosat} observations. Due to the fast optics of the XRT 
(Tr\"umper 1983) and the high sensitivity of the PSPC instrument, it was found 
(Snowden et al.\ 1991) that
the X-ray intensity, $I_x$, of a line of sight passing through the Draco nebula
was substantially attenuated. Specifically, a satisfactory fit for the 1/4 keV band 
count rate was obtained by a simple extinction law,
$I_x = I_f + I_b \exp[-\sigma(N_{\rm H})N_{\rm H}]$, with $I_f$ and $I_b$ 
denoting the foreground and the background
intensity and $\sigma(N_{\rm H})$ the H\,{\sc i} absorption cross section, 
respectively. Accordingly, roughly 50\% of the emission is from beyond
the Draco cloud; with distance limits between 300--1500 pc, this was clearly
{\it outside} the Local Hot Bubble, and thus in contradiction with the standard 
assumption of this model.
A more recent account of the 1/4 keV emission and its spatial (both local 
and halo) distribution is given by Snowden et al. (1998). 
We shall show here that in particular local winds push highly ionized gas into 
the halo. This gas cools but remains highly ionized. Its spectral
signature is consistent with observations in the Wisconsin Survey and with 
a {\sc Rosat} PSPC pointing toward the north Galactic
pole .   

The first search for diffuse soft X-ray emission in external galaxies was
undertaken by Bregman \& Glassgold (1982) and led essentially to a
negative result. A few years later, a marginal detection of an X-ray halo,
associated with the highly inclined spiral galaxy NGC$\,$4631, was
reported by Fabbiano \& Trinchieri (1987) with the {\sc Einstein}
satellite. More recently, extended diffuse soft X-ray emission of
another edge-on spiral galaxy, NGC$\,$891, was detected with {\sc Rosat}
(Bregman \& Pildis 1994), with a scale height of $\sim 2.4 \, {\rm kpc}$, 
comparable to the
scale height of the diffuse ${\rm H}{\alpha}$-emission (Dettmar 1992).
However, owing to its low Galactic latitude, the Galactic foreground 
absorption is large, and hence the measured count rate was too low to 
discriminate between different temperatures in the diffuse emission. 
Carrying out deep {\sc Rosat} PSPC observations of NGC$\,$4631, Wang et 
al.\ (1995) presented spatially separated spectra for the disk and the halo 
of NGC$\,$4631 in the energy range between 0.15 and 2.0$\,$keV. They found
extended, diffuse emission in the 1/4 keV band 
reaching out to more than 8 kpc above the galactic midplane. There was
also evidence for diffuse X-rays in the harder energy band (0.5--2.0 keV)
with about half the vertical extension of the 1/4 keV band. From the spectral
analysis, using a Raymond \& Smith (1977) fit model, Wang et al.\ (1995) 
concluded
that after correction for foreground H\,{\sc i}-absorption, a single
temperature fit of the halo emission is incorrect and that at least a
two-temperature fit is required. However, these authors emphasized that 
``... in reality there is probably a continuous temperature distribution 
in the halo gas.''
We expect spiral galaxies like NGC$\,$891 and NGC$\,$4631 
to show a similar dynamical and thermal behaviour of their halo plasmas 
than our Galaxy.  
NGC$\,$891 is often referred to as a twin galaxy to the Milky Way 
(cf.\ Dettmar, 1992), but NGC$\,$4631 is generally considered not to be a typical 
example for a normal spiral galaxy, because there is strong evidence 
for a gravitational disturbance in the past by a companion galaxy. 
However, the important 
physical process, responsible for an extended halo, is most likely the star 
formation rate (SFR), which drives an outflow, and 
which may have been somewhat enhanced in the past. At present the SFRs  
are the same in these two spiral galaxies within a factor of two; 
also the scale heights, both for the 
nonthermal radio halo and the diffuse ionized gas are comparable 
(cf.\ Dettmar 1992), thus corroborating our previous arguments. Moreover, 
more localized outflows have been reported 
for galaxies with enhanced SFR like M$\,82$ (e.g.\ Schaaf et al.\ 1989)
and NGC$\,$253 (Vogler \& Pietsch 1999; Pietsch et al.\ 1999) to which our 
theory also applies. 

In an earlier paper (Breitschwerdt \& Schmutzler 1994; henceforth Paper I) 
we have reported on how the soft X-ray background (SXRB) can be successfully 
modeled by the fast adiabatically expanding plasma in a galactic wind 
and in the Local Bubble. Here we elaborate on and extend these 
earlier investigations, showing in detail how the resulting non-equilibrium  
distribution of highly ionized species evolves in a galactic wind.
The plasma in the Local Bubble will be discussed in a separate paper. 
   
In Section~2 we describe in detail how dynamical and thermal evolution of the
ISM are intertwined, and how a successful method of a self-consistent
treatment works. In Sect.~3, the initial conditions and the physical state of the 
gas are discussed. Section~4 gives an overview of global and local galactic wind 
flows and their dynamical and thermal characteristics.
Corresponding X-ray spectra are presented in Section~5 as well as implications 
for the determination of interstellar temperatures from line ratios. In 
Sect.~6, we discuss and summarize the main results of our investigations.

\section{Self-consistent dynamical and thermal evolution of the ISM}
Outflows and related dynamical processes have been calculated by numerous
authors in different contexts, e.g.\ outflows from young stellar objects, 
from stellar wind bubbles, SNRs and from galaxies. Here we 
specifically refer to 
disk-halo related outflows like the expansion of superbubbles into 
a density stratified ambient medium (e.g.\ Tomisaka 1991), the flow 
obtained from a Galactic fountain (Bregman 1980; Kahn 1981; Avillez et al., 
1998) or, emphasizing superbubbles as the underlying sources, from chimneys 
(Norman \& Ikeuchi\ 1989).  For instance, the work by
Wang et al.\ (1995) and by Tomisaka \& Bregman (1993) addresses the dynamics of
galactic outflows (winds) and their spectral signatures. However, their 
approaches assume a simplified treatment of radiative cooling. More 
specifically, Tomisaka \& Bregman (1993) tried
to model the widely extended diffuse X-ray emission around M$\,$82. They
calculated a thermal galactic wind expanding into a static gas halo using
a 2-D hydrodynamical code. However, the radiative cooling term in their
calculation is based on the assumption of collisional ionization 
equilibrium (CIE).
It is only for comparison with the observed X-ray spectra that they discuss their
hydrodynamical model in terms of time-dependent ionization. Their arguments 
are mainly based on results of numerical models for young SNRs obtained by 
Hamilton et al.\ (1983).

Suchkov et al.\ (1994) have described in some detail the geometry and
structure of galactic superwinds interacting with disk and halo gas.
These authors also derive X-ray spectra and argue that the bulk of the 
soft X-ray emission should originate from the shocked disk and halo 
material, whereas the shocked wind would emit at significantly  
higher temperatures. But again these conclusions are based on CIE using 
a standard Raymond \& Smith (1977) cooling function.  
However, they and other authors obviously didn't realize the importance of 
the strong coupling between dynamics, thermal energetics and ionization states.
In particular in the case of superwinds, fast adiabatic expansion drives 
the system rapidly into non-equilibrium as we will show below. 

\subsection{The basic equations of the dynamics}
The dynamics of CR driven galactic winds have been discussed
comprehensively by Breitschwerdt et al.\ (1991). The description of
wind flows employing a three fluid model is based on the overall
conservation laws of mass, momentum and energy for the thermal plasma,
mean magnetic field, CRs and MHD wave field together with the hydrodynamic
equivalent of a CR transport, and a wave energy exchange equation.

In their general form these conservation laws can be written as
\beq
\frac{\partial\varrho}{\partial t} + \nabla (\varrho \vec{u})  =  q
\label{dyn-eq-1}
\eeq
\beq
\frac{\partial}{\partial t} (\varrho \vec{u}) +
\nabla \hbox{\cbf T} = \varrho \vec{F} + \: \vec{m}
\label{dyn-eq-2}
\eeq
\beq
\frac{\partial W}{\partial t} + \nabla \vec{S}  =  \varrho \vec{u}
(\vec{F} + \vec{m}) + \: \hbox{\cal E}\:,
\label{dyn-eq-3}
\eeq
where $q$, $\vec{m}$, {\cal E} denote sources and/or
sinks of mass, momentum and energy, respectively, and by $\vec{F}$ an external 
body force  (e.g.\ gravity) is specified. 
{\cbf T}\, is the momentum flux density tensor, $W$ the total energy 
density and $\vec{S}$ the energy flux density for the system as a whole, given by
\beq
\hbox{\cbf T}=\varrho\vec{u}\otimes\vec{u}+\left[\pg + \pcr + \frac{\langle
(\delta\vec{B})^2\rangle}{8\pi} + \frac{B^2}{8\pi}\right]\,
\cdot\,\hbox{\cbf I} - \frac{\vec{B}\otimes\vec{B}}{4\pi}
\label{dyn-eq-4}
\eeq
\beq
W=\frac{1}{2}\varrho u^2 + \frac{\pg}{\gag -1} + \frac{\pcr}{\gc -1} +
\frac{\langle (\delta\vec{B})^2\rangle}{4\pi} + \frac{B^2}{8\pi}
\label{dyn-eq-5}
\eeq
\beqarn
\vec{S}&=&\left(\frac{1}{2} u^2 + \frac{\gag}{\gag - 1}
\frac{\pg}{\varrho}\right)\varrho\vec{u} + \frac{1}{\gc - 1}
\Bigl[\gc\pcr\left(\vec{u} +\vec{v}_{\rm A}\right) \nonumber \\
& & -\bar\kappa\nabla\pcr\Bigl] +
\frac{\langle(\delta\vec{B})^2\rangle}{4\pi}
\left[\frac{3}{2}\vec{u} + \vec{v}_{\rm A}\right] + \frac{\vec{E}\times
\vec{B}}{4\pi}\:.
\label{dyn-eq-6}
\eeqarn
Here we describe the thermal plasma by its mass density $\varrho$, thermal
pressure $\pg$, velocity $\vec u$ and adiabatic index $\gag$; we treat the
CRs hydrodynamically through their pressure 
$\pcr = {4 \pi \over 3} \int_{0}^{\infty} dp \, w \, p^3 \,
f(\vec{x}, p, t)$ (with $f(\vec{x}, p, t)$ being the isotropic part of the 
particle distribution function in phase space, and $w$ and $p$ denote the 
particle speed and momentum, respectively), energy density
$\pcr/(\gc -1)$ with $\gc = 4/3$ (5/3) for ultra-relativistic (non-relativistic)
particles, and diffusive energy flux density $-\bar\kappa\nabla\pcr/(\gc -1)$,
where $\bar\kappa$ denotes the (Rosseland) mean CR diffusion coefficient.
The electromagnetic effects of the mean fields are considered by the Maxwell
stresses in Eq.~(\ref{dyn-eq-4}) and the Poynting flux vector
$(\vec{E}\times\vec{B})/4\pi$ in Eq.~(\ref{dyn-eq-6}), where $\vec{E}$ is the
electric and $\vec{B}$ the magnetic field. 

$\vec{E}$ and $\vec{B}$ are derived from Maxwell's equations, which under the 
assumption of ideal MHD read
\beq
\vec{E} = -\frac{1}{c} \, (\vec{u} \times \vec{B})\:,
\label{dyn-eq-7}
\eeq
and
\beq
\frac{\partial \vec{B}}{\partial t} = -c \, (\nabla \times \vec{E}) \:.
\label{dyn-eq-8}
\eeq
It is worth noting that the requirement of the magnetic field being free 
of divergence is contained in Faraday's law as an initial condition.

The wave field is resonantly generated through a so-called streaming 
instability (Lerche 1967; Kulrud \& Pearce 1969), which arises from a
small scale anisotropy in the pitch angle distribution of the CRs due to 
a spatial pressure gradient $\nabla\pcr$ of the CRs streaming away from the 
Galaxy. MHD (or for simplicity Alfv\'enic) waves satisfying a 
gyro-resonance condition are most efficiently excited (e.g.\ McKenzie
\& V\"olk 1982).
The Alfv\'en velocity is denoted by $\vec{v}_{\rm _A}$ and the mean square
fluctuating magnetic field amplitude by $\langle (\delta\vec{B})^2\rangle$.
The symbols $\otimes$ and {\cbf I} are used for the tensor product 
and the unit tensor, respectively. 

The CR transport equation describes the effects of convection and diffusion of
CRs in a scattering medium:

\beqarn
\frac{\partial}{\partial t}\left(\frac{\pcr}{\gc -1}\right) &+& \nabla\left\{
\frac{\gc}{\gc -1}(\vec{u} + \vec{v}_{\rm A})\pcr - \frac{\bar\kappa}{\gc -1}
\nabla\pcr\right\} \nonumber\\
&=& (\vec{u} + \vec{v}_{\rm A})\nabla\pcr + Q\:,
\label{dyn-eq-9}
\eeqarn

where the scatterers (Alfv\'en waves) move predominantly in the forward 
direction at the
Alf\'ven velocity $\vec{v}_{\rm A}$ relative to the plasma flow velocity
$\vec{u}$, and $Q$ may represent energy losses or gains other than those due to
adiabatic volume changes and resonant wave generation by the 
streaming instability. 

The wave energy exchange equation describes the effects of the background flow
and the CRs on the waves:
\beqarn
2\, \frac{\partial}{\partial t} \pw &+&
\nabla\left\{\pw \left[3 \, \vec{u} + 2\, \vec{v}_{\rm A}\right]\right\} 
\nonumber \\
&=&\vec{u}\nabla\pw - \vec{v}_{\rm A}\nabla\pcr + L\:,
\label{dyn-eq-10}
\eeqarn
where the wave pressure is defined by
\beq
\pw = \frac{\langle(\delta\vec{B})^2\rangle}{8\pi} \:,
\label{dyn-eq-11}
\eeq
and $L$ denotes additional wave energy losses or gains. 
It has been shown for global winds with a total plasma beta  
$\beta_{\rm tot} = 8\pi (\pg + \pcr)/B^2 > 0.1$ that nonsaturated
nonlinear Landau 
damping may dominate the advection of waves in the plasma and hence lead to 
local dissipative heating (Zirakashvili et al.\ 1996; Ptuskin et al.\ 1997). 
Although in general such a redistribution of energy between the fluid components 
will change the dynamics and the resulting emission spectra, this effect has 
not been taken into account in this paper, but will be discussed elsewhere.   

The system is closed by
the definition of the Alfv\'en velocity
\beq
\vec{v}_{\rm A} = \sqrt{\frac{|\vec{B}|}{4\pi\varrho}}\frac{\vec{B}}{|\vec{B}|}
\:.
\label{dyn-eq-12}
\eeq

A more detailed discussion of the equations and their special terms is given in
Appendix A of Breitschwerdt et al.\ (1991) and in Breitschwerdt (1994).

\subsection{Simplified model equations for the dynamics} 
Using the concept of
flux tubes with cross section area $A(s)$, where in the following all
the variables are taken to be functions of the streamline coordinate
$s$ only, we now concentrate on the system of dynamical equations which we have
to solve for our special models. Here we neglect any source or loss
term ($q=Q=L=0$ as well as $\vec{m}=0$) other than radiative cooling 
({\cal L}) and heating ({\cal G}), and assume the case of strong scattering 
of CRs, i.e. the averaged diffusion coefficient vanishes ($\bar\kappa=0$). 
Thus the net energy loss term is given by 
\beq
\hbox{\cal E} = \hbox{\cal G} - \hbox{\cal L}
\label{dyn-eq-13} \,.
\eeq
The equations of continuity and absence of magnetic monopoles read
\beq 
\varrho \, u \, A = const. \,, 
\label{dyn-eq-13a} 
\eeq
and 
\beq 
B \, A = const. \,,
\label{dyn-eq-13b}
\eeq
respectively. 
The ``heat equation" takes the form
\beq
\frac{d\pg}{ds}=\cg^2\frac{d\varrho}{ds}+(\gag-1)\frac{\hbox{\cal E}}{u} 
\label{dyn-eq-14}
\eeq
and the wave energy exchange and CR transport are described by
\beq
\frac{d\pw}{ds}=\frac{(3u+\va)}{2(u+\va)}\frac{\pw}{\varrho}\frac{d\varrho}{ds}
-\frac{\va}{2(u+\va)}\frac{d\pcr}{ds} \,,
\label{dyn-eq-15}
\eeq
\beq
\frac{d\pcr}{ds}=\frac{\gc\pcr}{(u+\va)}\left[\left(u+\frac{\va}{2}\right)
\frac{1}{\varrho}\frac{d\varrho}{ds}\right]\:.
\label{dyn-eq-16}
\eeq
Under the above assumptions, two adiabatic integrals can be readily
derived. One is the wave action integral and is given by
\beq
A\left[2\pw\frac{(u+\va)^2}{\va}+\frac{(u+\va)\gc\pcr}{\gc-1}\right]=
{\rm const.} \,,
\label{dyn-eq-17}
\eeq
and the other one results from the CR transport equation: 
\beq
\pcr [(u+\va)A]^{\gc}={\rm const.} \,.
\label{dyn-eq-18}
\eeq
With the help of the other equations, the momentum equation can be transformed 
into a ``wind equation''
\beq
\frac{du}{ds}=
\frac{u\left[c^2_*(A'/A)-g_{\rm eff}+\frac{(\gag-1)}{\varrho u}\hbox{\cal E}
\right]}{u^2-c^2_*}\:;
\label{dyn-eq-19}
\eeq
here the ``compound sound speed'' $c_*$ corresponds to the long wavelength limit 
obtained from the kinetic equations,
\beqarn
c^2_*&=&\gag\frac{\pg}{\varrho}+\gc\frac{\pcr}{\varrho}\frac{(M_{\rm A}+
\frac{1}{2})^2}{(M_{\rm A}+1)^2}+\frac{\pw}{\rho}
\frac{3M_{\rm A}+1}{2(M_{\rm A}+1)}\cr
&=& \cg^2+c_{\rm C}^2+c_{\rm W}^2\:,
\label{dyn-eq-20}
\eeqarn
where $\cg$, $c_{\rm C}$ and $c_{\rm W}$ denote the ``sound speeds'' 
for gas, CRs and waves, respectively, 
and $g_{\rm eff}$ describes the effective gravitational acceleration, derived
from a mass distribution for the Galaxy, consisting of a bulge, disk and 
dark matter halo component (see Breit\-schwerdt et al.\ 1991). The
Alfv\'en Mach number is defined as $M_{\rm A}=u/\va$.

The dynamics are calculated assuming an ultrarelativistic CR component 
($\gc = 4/3$) and $\gag=5/3$. 
Although the latter assumption is not consistent with the thermal treatment of
the gas where
the variation of $\gag$ is implicitly taken into account (cf.\ Sect.\ 2.3 and
Schmutzler \& Tscharnuter 1993), this formal inconsistency is neutralized, 
because the internal energy of the gas is balanced in the thermal equations 
and because we iterate each system of equations by using the updated
quantities of the other one.

\subsection{The thermal processes}
We use a net radiative cooling rate which is a complicated function of several
physical processes. They depend on gas temperature, gas density, element
abundances, ionization states as well as the spectral energy density 
$U_{\nu}$ of external illuminating continuum photons. 
For the detailed calculations we use the code
HOTGAS, which has been described in Schmutzler \& Tscharnuter (1993). Here we 
just summarize the relevant physics in a short list of the processes taken into
account and give the references for the atomic data.

The ten most abundant elements H, He, C, N, O, Ne, Mg, Si, S and Fe are
considered with cosmic abundances given by Allen (1973). Ionization is driven by
collisions with both thermal electrons and neutral atoms (Hollenbach \& McKee 
1989), as well as by charge exchange and photoionization. Hydrogen
and helium can also be ionized by suprathermal electrons. These are produced by
high energy photoionization and Auger effect (Shull 1979; Halpern \& Grindlay
1980; Binette et al.\ 1985). We use the rates for thermal ionization including
excitation-autoionization and charge exchange given in the excellent compilation
of Arnaud \& Rothenflug (1985) and the revision for iron according to the paper
of Arnaud \& Raymond (1992). In addition to charge exchange, which is also a
recombination process, we consider both types of direct recombination due to
two-body collisions: radiative and dielectronic recombination, where the last
one is the inverse process to excitation-autoionization and is considered as a
density dependent process (Jordan 1969). In the case of singly ionized atoms
the code HOTGAS also takes corrections for three-body recombination into 
account, which
is the inverse process to collisional ionization (Hollenbach \& McKee 1989).
The radiative recombination coefficients used here include all possible
recombinations into excited levels and into the ground level. For hydrogen yet,
we apply the ``on-the-spot"-approximation (Osterbrock 1974). The coefficients
are taken from Tarter (1971), Shull \& Van Steenberg (1982a, 1988b), Aldrovandi
\& P\'equignot (1973, 1976), Arnaud \& Rothenflug (1985) and Arnaud \& Raymond
(1992). In addition, we assume that the Lyman continuum photons of recombining 
helium, the second most abundant element, ionize hydrogen.

The energy input from the external radiation field is taken into account due to
the most important interactions of photons with atoms, ions and electrons. The
total photoionization cross sections are taken from Reilman \& Manson (1978,
1979) and are interpolated as a series of power laws. The required subshell
cross sections are derived from the total one. For hydrogen and hydrogen like
ions in the {1-s} state we use the exact cross section formula (e.g., Vogel
1972). If the photon energies are much higher than the ionization potential, the
extrapolated cross sections fall below the Thomson cross section. In these cases
we follow the arguments of Halpern \& Grindlay (1980) and calculate the Compton
ionization cross sections for all ions.

High energy photons may ionize more likely an inner shell than an outer one. If
the following reconfiguration of electrons releases a sufficient amount of 
energy,  
there is a certain probability of an additional emission of one or more of the
outer electrons (Auger effect). We use the probabilities and number of
emitted electrons per inner shell ionization worked out by Weisheit (1974).
 
The net cooling rate is balanced in detail by the appropriate heating terms due
to photoionization, Compton ionization, Auger effect and the net energy exchange
rate between photons and electrons due to Compton scattering (Levich \&
Sunyaev 1970, 1971) as well as by the most important radiative energy-losses. We
determine the energy-loss due to Coulomb collisions of electrons and ions using
the formula of thermal bremsstrahlung (Novikov \& Thorne 1973) with the
frequency averaged Gaunt factor given by Karzas \& Latter (1961). Both types of
recombination transform thermal energy into radiative loss. We derived the
energy-loss rates from the corresponding volume emissivity given by Cox \&
Tucker (1969). In case of dielectronic recombination the first excitation
energies tabulated in Landini \& Fossi (1971) are used. 
However, the most important
energy-loss due to thermal particle interactions results from collisional
excitation followed by spontaneous line emission including two-photon continuum
emission. Two-photon emission and some special transitions, such as fine
structure lines, semiforbidden and forbidden lines have been implemented in a
density dependent form (Mewe et al.\ 1985; Innes 1992). In total we consider
1156 line transitions in the spectral range from 1\AA\ to 610$\,\mu$m. The data
are based on the work of Kato (1976), Stern et al.\ (1978), Osterbrock (1963,
1971), Osterbrock \& Wallace (1977), Jura \& Dalgarno (1972), Penston (1970),
Giovarnadi et al.\ (1987) and Giovarnadi \& Palla (1989).

In our calculations only tenuous gases ($n \ll 1\,\rm cm^{-3}$) are  
considered and therefore
neither three body recombination nor the density dependent suppression of some of 
the thermal processes contribute.

\subsection{The equations of a thermally self-consistent approach}
The set of equations to be solved consists of energy balance, time dependent 
state of ionization, charge conservation, and the equation of state. 
In addition, we have
to provide the time variations of the pressure (or density or temperature) at
least in a global way. Considering the 10 most abundant elements, the system of
equations contains 103 ordinary differential equations together with 14
algebraic relations. We note that an adequate description of a pure hydrogen
plasma can be given by 2 ordinary differential equations and 5
algebraic relations. In any case, such a system of non-linear ordinary
differential equations is well known to be stiff and therefore requires an 
implicit method for an efficient solution (Schmutzler \& Tscharnuter 1993). 

We use the variables gas temperature $T$, pressure $P$, mass density $\varrho$,
number of electrons per unit mass $n_{\rm e}/ \varrho$, number of ions per
unit mass $n_{Z,z}/\varrho$ and internal energy per unit mass $U$; here $Z$
and $z$ denote the nuclear and effective charge of an ion, respectively. 
The ions and electrons are assumed to have the same Maxwellian 
temperature, because the equilibrium distribution is typically reached on time 
scales shorter than recombination or ionization processes or radiative 
losses do occur. 

The equation of energy balance reads:
\beq
{dU\over dt} - {P\over\varrho}\,{d\ln\varrho\over dt} +
{{\hbox{\cal L} -  \hbox{\cal G}}\over\varrho}=0\:,
\label{therm-eq-1}
\eeq
where $U$ is the specific internal energy, defined as
\beq
U - \sum_{Z}\sum_{z=1}^{Z}\left({n_{Z,z}\over \varrho}
\sum_{z'=0}^{z-1}I_{Z,z'}\right) - {3\over 2}{P\over\varrho}=0\:.
\label{therm-eq-2}
\eeq
Here $I_{Z,z'}$ gives the ionization potential of an ion with nuclear charge
$Z$ and effective charge $z'$. This definition implicitly considers the fact
that the commonly used ratio of specific heats $\gamma=c_P/c_V$ is not a
constant (c.f. Schmutz\-ler \& Tscharnuter 1993). The net radiative cooling rate
{\cal L} -- {\cal G}, expressed by the cooling function {\cal L} and a possible
heating function {\cal G}, represents only that part of the total radiative
energy-loss which directly couples to the velocity distribution of the electrons,
i.e. the thermal evolution of the gas. It depends on temperature, electron density
and ionization states. Moreover, the contribution of an external photon field
depends on photon density and energy spectrum.

Ion and electron densities are determined by the ionization and recombination
rates, which also depend on temperature, particle densities and external photon
field. Thus we have to solve at least 1 (in the case of a pure hydrogen plasma)
or 102 (in case of a plasma consisting of the 10 most abundant elements)
ordinary differential equations balancing the ionization states of all ions with
$z\not=0$ of the form:
\beq
-{d\left({n_{Z,z}\over \varrho}\right)\over dt} + {1\over \varrho}\,F_{Z,z}
(T,n_{\rm e},n_{Z,z},\tilde n_{\tilde Z,\tilde z},
\zeta^{\rm phot}_{Z,z})=0\:.
\label{therm-eq-3}
\eeq
The functions $F_{Z,z}$ contain all ionization and recombination rates which,
due to charge exchange and ionization by collisions of neutral atoms, may also
depend on the particle density of other ions $\tilde n_{\tilde Z,\tilde z}$.
The dependence on possibly interacting photons is indicated by
$\zeta^{\rm phot}_{Z,z}$.

The abundances of atoms ($z=0$) are then given by the (1 to) 10 equations, in which
also the relative chemical abundances $X_{Z}$ in a gas with changing mass
density are specified:
\beq
\sum_{z}{n_{Z,z}\over \varrho} - {X_{Z}\over \langle\mu\rangle}=0\:,
\label{therm-eq-4}
\eeq
where $\langle\mu\rangle$ denotes a normalized mean mass for a given chemical
composition.

Charge conservation determines the electron density:
\beq
\sum_{Z,z}{n_{Z,z}\over \varrho}\,z - {n_{\rm e}\over \varrho}=0\:.
\label{therm-eq-5}
\eeq

Pressure, temperature and mass density are tied into the equation of state by:
\beq
P - \varrho\left(\sum_{Z,z}{n_{Z,z}\over \varrho} + {n_{\rm e}\over \varrho}
\right)\,k\,T=0\:.
\label{therm-eq-6}
\eeq

In order to take into account the dynamics in a global way, we have to fix
one additional relation, for example, the time variation of the density (or
pressure or temperature) according to the time scales of the dynamical behaviour 
of the gas: 
\beq
\varrho - f(t)=0\:.
\label{therm-eq-7}
\eeq

A coupling between both systems of equations can be achieved by 
Eq.~(\ref{therm-eq-7}) on the thermal, and by the net radiative cooling term
$(\hbox{\cal L} -  \hbox{\cal G})/\varrho$ in Eq.~(\ref{therm-eq-1}) on the 
dynamical side. Although the 
dynamical equations in the stationary case do not contain an explicit time scale
one can easily find the flow time
\beq
\tau_{\rm flow}(s)=\int \limits_{s_0}^s {d\,s'\over u(s')}\:,
\eeq
which allows to transform the dynamically determined density $\varrho(s)$ into a
time-dependent function $\varrho(\tau)$ as input for Eq.~(\ref{therm-eq-7}). The
solution of the thermal equations by Newton-Raphson iteration provides the net
radiative cooling term $(\hbox{\cal L} -  \hbox{\cal G})/\varrho$ as function of
$T$ or (and) $s$. By using this solution as an input for
the dynamical equations, a new
time-dependent density can be found. Since $\varrho(\tau)$ as well as
$(\hbox{\cal L} -  \hbox{\cal G})/\varrho$ are determined in a discrete form and
each system of equations defines its appropriate step widths, we use taut cubic
splines for interpolation. The complete procedure, which is shown in the 
form of a flow chart in Fig.~\ref{dia1}, has to be iterated until the
temperatures of each system, $T_{\rm d}(s)$ and $T_{\rm t}(s(t))$, converge.
The comparison of the two temperatures provides a good convergence check, 
because temperature is not used as a primary variable, and it 
is calculated independently in each system of equations.
Typically one needs 3--4 iterations to find an excellent agreement between
both temperatures. The deviations are found to be less than 0.5\% for distances
$s$ between the base of the flux tube and the critical point, and they are less
than 3\% at distances greater than 10 times that of the critical point.

The self-consistent solution describes the time-dependent (here this is
equivalent to space-dependent) state of the gas in a volume element flowing
along the flux tube. 
As a consequence of stationary wind models the state of 
the gas varies as a function of distance but it is constant at any given 
position.
%
%
\renewcommand{\textfraction}{0.5}
\begin{figure*}[htbp]
\unitlength1cm
\begin{picture}(14.5,19.5)
%
%
\put(0,9.5){\unitlength1cm
\begin{picture}(14.0,10.0)
\thicklines \put(0,0){\framebox(14.0,10.0){}} \thinlines
\put(7.0,9.7){\makebox(0,0){\bf Dynamic Equations}}
\put(5.0,7.5){\unitlength1mm\begin{picture}(40,15)
    \put(0,0){\framebox(40,15){Mass Conservation}}
    \put(19,2){$\rho$} \end{picture}}
\put(5.0,4.0){\unitlength1mm\begin{picture}(40,15)
    \put(0,0){\framebox(40,15){Momentum Conservation}}
    \put(19,2){$\vec{u}$}
    \put(20,15){\vector(0,1){20}}
    \put(20,35){\vector(0,-1){20}}
    \put(40,8){\vector(1,0){7}}
    \put(47,8){\vector(-1,0){7}}
    \put(20,0){\vector(0,-1){20}}
    \put(20,-20){\vector(0,1){20}} \end{picture}}
\put(0.3,4.0){\unitlength1mm\begin{picture}(40,15)
    \put(0,0){\framebox(40,15){CR Transport}}
    \put(19,2){$\pcr$}
    \put(23,0){\vector(2,-1){40}}
    \put(63,-20){\vector(-2,1){40}}
    \put(40,8){\vector(1,0){7}}
    \put(47,8){\vector(-1,0){7}}\end{picture}}
\put(9.7,4.0){\unitlength1mm\begin{picture}(40,15)
    \put(0,0){\framebox(40,15){Maxwell-Equations}}
    \put(19,2){$\vec{B}$} \end{picture}}
\put(5.0,0.5){\unitlength1mm\begin{picture}(40,15)
    \put(0,0){\framebox(40,15){Energy Conservation}}
    \put(19,2){$\pg$}
    \put(23,15){\vector(2,1){40}}
    \put(63,35){\vector(-2,-1){40}} \end{picture}}
\put(0.3,0.5){\unitlength1mm\begin{picture}(40,15)
    \put(0,0){\framebox(40,15){}}
    \put(20,13){\makebox(0,0)[t]{Wave Energy}}
    \put(20,9){\makebox(0,0)[t]{Exchange Equation}}
    \put(20,5){\makebox(0,0)[t]{$\pw$}}
    \put(20,15){\vector(0,1){20}}
    \put(20,35){\vector(0,-1){20}}
    \put(23,15){\vector(2,1){40}}
    \put(63,35){\vector(-2,-1){40}}
    \put(40,8){\vector(1,0){7}}
    \put(47,8){\vector(-1,0){7}}\end{picture}}
\end{picture}}
%
%
\newsavebox{\bedzweig}
\savebox{\bedzweig}(0,0)[bl]{
\thicklines
\put(0,0.5) {\line(-2,-1){1.0}}
\put(0,0.5) {\line(2,-1){1.0}}
\put(0,-0.5) {\line(-2,1){1.0}}
\put(0,-0.5) {\line(2,1){1.0}}
\put(0,0.5) {\line(0,1){0.5}}
\put(0,-0.5) {\line(0,-1){0.5}}
\put(1.0,0.0) {\line(1,0){1.0}}
\put(-1.0,0.0) {\line(-1,0){1.0}}
\put(-1.1,0.1){\makebox(0,0)[br]{no}}
\put(1.1,0.1){\makebox(0,0)[bl]{yes}}  }
\put(0.0,6.5){\unitlength1cm
\begin{picture}(14.0,3.0)
\thicklines
\put(7.0,1.5){\usebox{\bedzweig}\makebox(0,0.0){$T_d=T_t$}}
\put(10.5,1.5){\oval(3.0,1.0)}
\put(10.0,1.4){\bf STOP}
\put(5.0,1.5){\vector(0,1){1.5}}
\put(5.0,1.5){\vector(0,-1){1.5}}
\put(7.0,3.0) {\vector(0,-1){1.0}}
\put(7.0,0.0) {\vector(0,1){1.0}}
\put(7.3,2.5){\makebox(0,0){$T_d$}}
\put(7.3,0.5){\makebox(0,0){$T_t$}}
\put(4.4,2.0){\makebox(0,0){$\Gamma - \Lambda$}}
\put(4.1,1.0){\makebox(0,0){$\rho$, $\vec{u}$, $\tau_{\rm dyn}$}}
\end{picture}}
%
%
\put(0.0,0.0){\unitlength1cm
\begin{picture}(14.0,6.5)
\thicklines \put(0,0){\framebox(14.0,6.5){}} \thinlines
\put(7.0,6.2){\makebox(0,0){\bf Thermodynamic Equations}}
\put(5.0,4.0){\unitlength1mm\begin{picture}(40,15)
    \put(0,0){\framebox(40,15){Internal Energy Balance}}
    \put(19,2){$U$}
    \put(40,8){\vector(1,0){7}}
    \put(47,8){\vector(-1,0){7}} \end{picture}}
\put(0.3,4.0){\unitlength1mm\begin{picture}(40,15)
    \put(0,0){\framebox(40,15){Ionization Balance}}
    \put(17,2){$n_{Z,z}$}
    \put(17,0){\vector(1,-1){20}}
    \put(37,-20){\vector(-1,1){20}}
    \put(40,8){\vector(1,0){7}}
    \put(47,8){\vector(-1,0){7}}\end{picture}}
\put(9.7,4.0){\unitlength1mm\begin{picture}(40,15)
    \put(0,0){\framebox(40,15){Equation of State}}
    \put(16,2){$\pg(\rho)$} \end{picture}}
\put(2.3,0.5){\unitlength1mm\begin{picture}(40,15)
    \put(0,0){\framebox(40,15){}}
    \put(20,13){\makebox(0,0)[t]{Chemical}}
    \put(20,9){\makebox(0,0)[t]{Abundances}}
    \put(20,4){\makebox(0,0)[t]{$X_Z$}}
    \put(23,15){\vector(1,1){20}}
    \put(43,35){\vector(-1,-1){20}}
    \put(40,8){\vector(1,0){14}}
    \put(54,8){\vector(-1,0){14}}\end{picture}}
\put(7.7,0.5){\unitlength1mm\begin{picture}(40,15)
    \put(0,0){\framebox(40,15){Charge Conservation}}
    \put(19,2){$n_{\rm e}$}
    \put(17,15){\vector(-1,1){20}}
    \put(-3,35){\vector(1,-1){20}}
    \put(23,15){\vector(1,1){20}}
    \put(43,35){\vector(-1,-1){20}} \end{picture}}
\end{picture}}
\end{picture}
\renewcommand{\textfraction}{}
\caption{Flow chart of a hydrodynamically and 
thermally self-consistent outflow, taken from Breitschwerdt (1994). 
The dynamical calculations provide $\rho$, $\protect\vec{u}$ and the 
dynamical time scale $\tau_{\rm dyn} = \int_{z_0}^{z}(dz/u)$, which 
serve as input values for the thermodynamic calculations. The resulting 
net cooling function $\tilde\Lambda = \Gamma - \Lambda$ is subsequently 
used for the following dynamical iteration. This procedure is repeated 
until the temperature of the dynamic calculation, $T_d$, and the 
temperature of the thermodynamic calculation, $T_t$, converge. }
\label{dia1} 
\end{figure*}
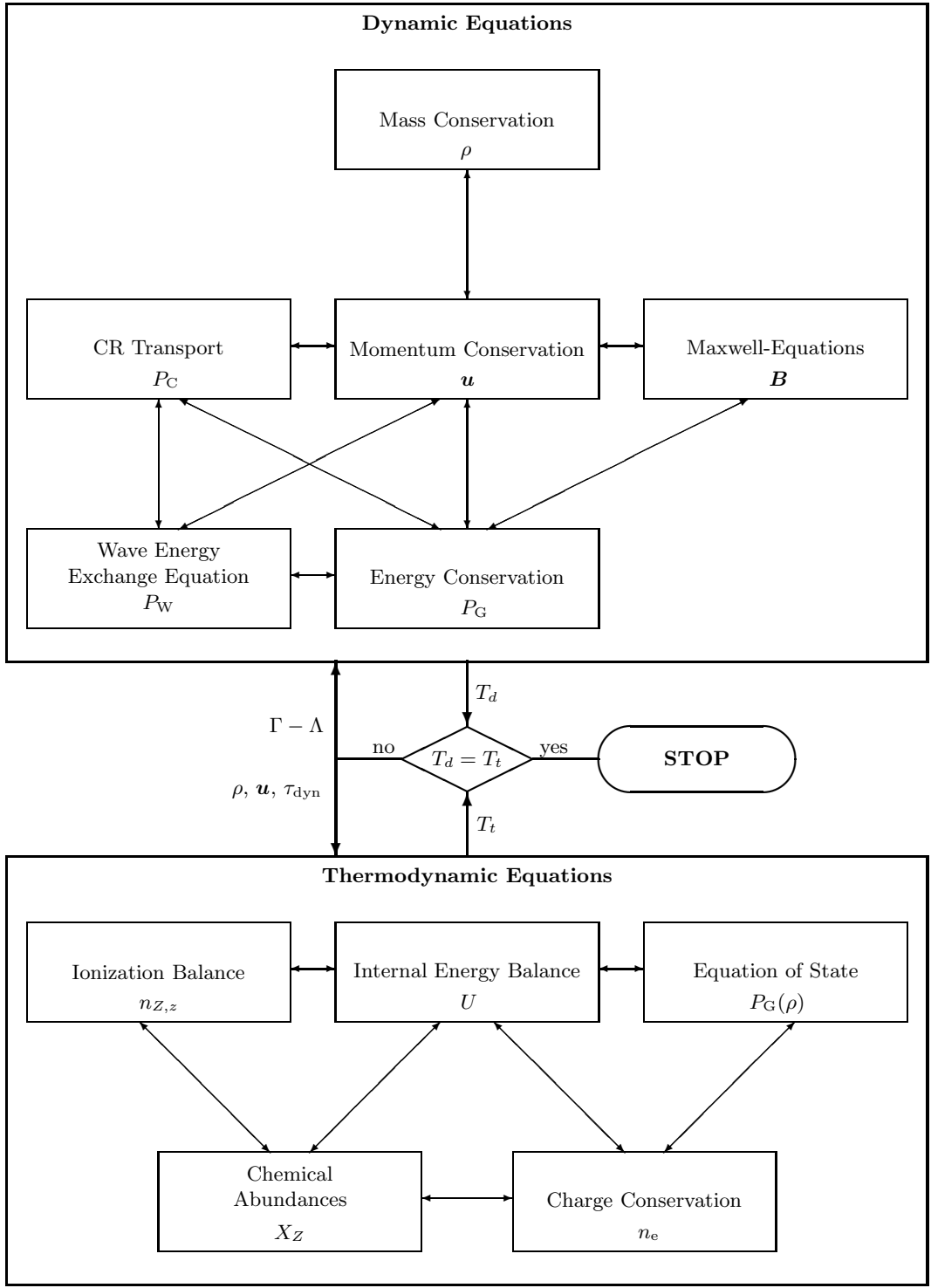

\section{The initial state of the gas}
The time-dependent evolution of a gas depends not only on the thermodynamic 
path but also on the state of the gas at a given time, the initial state. The 
gas state is determined by the temperature, the density, the chemical 
composition and the ionization states of the elements. In most cases the initial
state of the gas is unknown. Nevertheless, it should be possible to constrain the 
range of initial values by comparing observational results with those of 
calculated evolutions, if the latter ones are based on physically plausible 
assumptions.\par

The new generations of EUV and X-ray spectrometers, starting with EUVE, ASCA, 
AXAF and most importantly XMM, 
provide increasing resolution in energy and position. Thus, from a physical 
point of view, we expect the detection of many spectral features, which can 
only be understood  within the framework of time-dependent models of the HIM, 
like the ones we present here. Of 
course, the appropriate analysis of these observations takes much more effort
than the standard analysis of today, but we gain the chance of looking back in 
time and learning more about the ``history'' of the gas than by any fit with 
standard equilibrium models.

\subsection{Initial ionization states}

For galactic winds, driven by thermal gas and CRs, originating in SNRs or
superbubbles, one may distinguish three major cases concerning the initial
ionization states. 
The first one is the assumption of a gas just departing from 
collisional ionization equilibrium (CIE) or from an ionization state close to 
that. In CIE the ionization states are determined by the assumed initial
temperature. Slight modifications are produced, for instance, by photoionization
due to a locally acting photon field (e.$\,$g.\ the averaged stellar and
extragalactic photon field). Ionization states close to CIE may be realized,
if $T \geq 10^6\, {\rm K}$ and the gas is kept at comparable temperatures for a time 
longer than the relevant recombination time scales.
                     
Here we note that CIE in general is a bad approximation of the thermal state of plasmas 
with cosmic abundances for $T < 10^6\, {\rm K}$. Only plasmas with very low metallicities 
may approximately reach CIE, since then the radiative cooling time is longer
than the recombination time scales. Although heating processes reduce the net
cooling time as well, they directly affect the ionization stages. Thus the plasma may 
approach ionization equilibrium, but still differ from CIE.
\par

The second case concerning the initial ionization states can be understood as
follows: shock heated gas in SNRs may reach temperatures between $10^6\, {\rm K}$
and $10^8\, {\rm K}$. 
However, it is known that the life time of a 
single SNR may not be long enough for the heated gas ever reaching CIE. 
If energy between ions and electrons is transferred mainly by Coulomb collisions, 
the electrons would come into equilibrium with the ions on a time scale 
of the order of $\sim 5000 \, {\rm E}_{51}^{1/4}/n_0^{4/7} \, {\rm yr}$ 
(cf.\ Itoh 1978, Cox \& Anderson 1982); here ${\rm E}_{51}$ is the hydrodynamic 
energy released in a SN explosion, and $n_0$ is the number density of the 
ambient medium. Other collisional processes would 
operate on even longer time scales. This would only be significantly 
reduced, if heat conduction would be large (which is unlikely due to suppression 
by magnetic fields) or plasma instabilities would redistribute the energy more 
efficiently. In the case of non-equilibrium, a (local) galactic
wind, fed from such sources, starts with ``underionized'' gas 
(with respect to the ionization states given by CIE at that temperature), 
in which hydrogen already may be fully ionized, whereas the ionization of 
the heavier elements may have reached only the (energetically) lowest or at 
most intermediate levels.
\par

The third case is described by an ``overionized" gas. It can be produced, for
instance, by photoionization of gas in the neighbourhood of strong continuum 
sources, like X-ray binaries or an active galactic nucleus. Another possibility
is fast adiabatic cooling of hot ($T \sim 10^8\, {\rm K}$), almost completely
ionized gas expanding out of a superbubble. Because of recombination delay the
cooled gas of several $10^5$ or $10^6\, {\rm K}$ consists of many more highly
ionized species than a gas in CIE at these temperatures.

In this paper we concentrate on the first case. The other two cases are the subjects 
of further investigations.

\subsection{Initial density and temperature}
The particle density of the HIM is assumed to range from $10^{-2}$ to 
$10^{-4} \, {\rm cm}^{-3}$. The most plausible interval for the initial temperature 
of gas fed into a galactic wind ranges from a few times $10^5$ to a few times 
$10^6\, {\rm K}$. Both intervals, for temperature and density, cover those values 
derived from pressure equilibrium arguments in models of the ISM (e.g., McKee \&
Ostriker 1977). However, we are interested in dynamical processes in the ISM,
which require various phases of the ISM being not in pressure equilibrium, at least
in a local part. One finds that particle densities and temperatures in the given 
range provide a sufficient pressure to initiate a break-out of the Galactic disk.

The more moderate values of density and temperature may be typical for HIM 
distributed nearly uniform over the Galactic disk. As has been discussed by
Breitschwerdt et al.\ (1987, 1991) CRs are able to couple to the thermal gas and 
to drive a galactic wind. According to the global distribution of both, moderate HIM
and CRs, we expect global galactic winds.
 
The sources of the HIM, stellar wind bubbles, SNRs and superbubbles, produce 
gas at somewhat higher temperatures. At least in the vicinity of these sources 
the gas density and temperature reach values which are sufficient for a 
thermally driven break-out and wind. An extreme example of thermal winds is 
observed in M$\,82$ (e.g.\ Schaaf et al.\ 1989). 

Thus, as long as star formation is important for the evolution of the ISM,
we expect not only global galactic winds but also local winds to occur. The latter 
ones may start as thermally driven flows and proceed further out by the interaction 
with CRs. In Sect.~4 we present models for both types of galactic winds.

\subsection{The chemical composition}
The chemical composition of the gas in the Galactic plane is expected to be a
function of position. For local Galactic winds one may expect higher abundances
than solar, because of the enrichment by supernovae (SNe). However, it is 
unclear to what extent entrainment of ambient gas dilutes the mixture, as it 
is suggested by some ASCA observations (Ptak et al. 1997). 
In order to demonstrate the differences between
self-consistent calculations and those in which the thermal processes just
follow the dynamical time scale, we assume solar abundances for the models in 
this paper.
The analysis of future observations with high spectral resolution based         
on thermally and dynamically self-consistent models will help to determine the  
chemical abundances in our Galaxy as well as in other galaxies.                 

\subsection{The initial state of the nonthermal components of the ISM}
Among the nonthermal ISM components the CRs are the primary driving agents 
of a galactic wind in spiral galaxies. The waves are largely self-excited 
by the CR streaming, and the magnetic field, $\vec{B}$, is treated statically 
here; for the dynamical r\^ole of $\vec{B}$ we refer to Zirakashvili et al.\ 
(1996). The bulk of CRs below about $10^{15} \, {\rm eV}$ is most likely 
of Galactic origin and are believed to be generated by diffusive shock 
acceleration with an efficiency of up to 50\% 
of the available hydrodynamic SN energy (Berezhko \& V\"olk 1997).
CRs propagate diffusively through the ISM and their propagation is 
largely determined by collective effects. Treating them as a high energy gas is 
a short-cut that ignores the complicated dynamics resulting from the solution 
of a Fokker-Planck type transport equation. The energy density of the CRs 
is known from in situ measurements to be about about $0.5 \, {\rm eV}/{\rm cm}^3$ 
(s.~ V\"olk et al.\ 1989). Since CRs also fill the Galactic halo, and a 
particle reaching the solar system has spent considerable time out there, a value of 
this order should be representative for global winds. In the vicinity of 
SNRs and superbubbles, a higher value is possible. However, in order 
to be conservative, our initial value of the CR pressure, $\pcrn$, is a 
factor of 3 below the locally measured $0.5 \, {\rm eV}/{\rm cm}^3$ 
for global winds and only a factor of 2--3 higher for local winds. 

In the solar neighbourhood, energy equipartition between CRs and the magnetic 
field is roughly fulfilled. In our flux tube geometry, we describe the vertical 
component of $\vec{B}$ and have therefore assumed a magnitude which is a factor 
of 2--5 smaller than the average disk value of $|\vec{B}| \sim 5 \muG$. Since the 
wave field is predominantly generated by the CR streaming, we have assumed 
a negligible initial value of $\langle(\delta\vec{B})^2 \rangle/8 \pi$ of 
1\% of the regular magnetic field energy density.

\section{X-ray emission from galactic winds}
The discussion in the previous sections has emphasized the necessity to 
critically examine 
the dynamical and thermal history of the ISM, if one wishes to interpret 
observations. Although in some circumstances the convenient assumption
of CIE may be not too far off from the real situation,
there is always a possibility that in general such an approach
will be entirely misleading. Clearly, further and independent 
information is needed. For example a measurement of line widths, if 
possible, would give an upper limit for the temperature. However, in most 
cases, all that is available is an energy spectrum in a certain wave 
length range. Then, only a careful analysis of the physical state of the 
observed region and a determination of its dynamical and thermal 
properties will lead to a trustworthy interpretation.  
In the following, we shall demonstrate how the dynamical state of the ISM 
will change its spectral appearance in the EUV and X-ray wavebands.  

Galactic winds can be crudely divided into two classes: {\it global} 
and {\it local winds}. The winds that have been described in 
the literature were mostly global winds, arising from a global 
hot ISM, which cannot be trapped in the galactic potential well. The 
pressure forces that cause the gas to escape from a galaxy may simply 
be due to random thermal motions (e.g.\ Mathews \& Baker 1971), also
including  
centrifugal forces due to galactic rotation (Habe \& Ikeuchi 1980), or
they can also be supported by 
the nonthermal energy content of the CRs (e.g.\ Breit\-schwerdt et
al.\ 1987, 1991; Fichtner et al.\ 1991). A recent model by Zirakashvili et al.\
(1996) also takes into account the dynamical effects of magnetic stresses 
caused by a wound-up Parker type galactic magnetic field.
 
Star formation of massive hot stars takes place predominantly in 
OB associations. Since these stars evolve within less than 
$10^7$ years, SNRs and hypersonic stellar winds create a 
superbubble with a large overpressure compared to 
the average ISM. There is no way that such a region can 
be confined, e.g.\ by a global interstellar magnetic field, as it is assumed
in some  models (Edgar \& Cox 1993). Instead, an upward expansion of gas and
CRs will lead to elongated bubbles and chimneys. 
Some numerical simulations of superbubble expansion argue in favour of 
confinement within the thick extended H\,{\sc i} and/or H\,{\sc ii}-layer of the Galaxy 
(Tomisaka 1991; Mineshige et al.\ 1993), because of a magnetic field which is
parallel to the shock surface. Therefore, depending on the magnitude of 
the field, magnetic tension can in principle considerably decelerate the 
flow, so that it will eventually stall before break-out. However, such an 
idealized field configuration is unrealistic, because the Parker
instability (both linear and nonlinear) will break up the field into 
a substantial component parallel to the flow (Kamaya et al.\ 1996).
The presence of CRs and magnetic fields in galactic halos is well 
known by the observation of synchrotron radiation generated by the 
electronic component (e.g.\ reviews by Beck et al.\ 1996).
In our view, the combined overpressure of thermal gas and CRs will
therefore eventually 
drive a local mass outflow with a comparatively high speed, which is 
causally connected to one or more underlying superbubble regions. 
The mass that is transferred to the lower halo, however, exceeds the mass that 
is ultimately driven out to infinity, depending on the total thermal and CR 
energy available. Therefore a substantial fraction of the 
gas is expected to fall back onto the disk in a fountain or chimney type 
fashion. The interaction with uprising gas will create shear and thus add 
to the turbulence in the halo. We note parenthetically, that such a 
process will support any operating turbulent halo dynamo.   

The extreme case of star formation is realized in so-called starburst 
galaxies like M$\,$82 or NGC$\,$253, which are the most prominent and 
best studied examples. Conceptually, this just represents an 
extrapolation of local galactic winds to a region with enhanced star
formation rate (SFR), according to the observed high far-infrared luminosities 
(Rieke et al.\ 1980). In M$\,$82, radio observations (Kronberg et al.\
1985) show an emission region extending to about 600 pc along the major 
axis and about 100 pc in vertical direction, consistent with a SN rate 
of $0.1 - 0.3 \, {\rm yr}^{-1}$. This inevitably leads to a thermally 
driven galactic wind (Chevalier \& Clegg 1985), which has been 
directly observed in the form of an extended X-ray halo both with 
{\sc Einstein} (Fabbiano 1988) and {\sc Exosat} (Schaaf et al.\ 1989). 
Moreover, from
the extension of the observed radio continuum halo, it is inferred that 
the relativistic electrons must be advected by a galactic wind with an 
average speed between 2000 km/s (Seaquist et al.\ 1985) and 4000 km/s
(V\"olk et al.\ 1989), in order to compensate for heavy synchrotron and
Inverse-Compton losses. There is further support for this interpretation 
from multifrequency radio observations, which reveal a fairly flat 
spectral index (Seaquist \& Odegard 1991). Model calculations have
shown (Breitschwerdt 1994), that such a behaviour is naturally explained 
by an accelerating galactic wind flow, that compensates for losses 
increasing with vertical distance from the disk in a constant halo 
magnetic field. 
In NGC$\,$253 {\sc Rosat} PSPC and HRI observations have 
shown (Pietsch et al.\ 1999) both a nuclear and an extranuclear
outflow. It is interesting to note that the latter one, which is most 
prominent in an X-ray spur in the southern hemisphere, is causally 
connected to a region of increased star formation in the disk as 
seen in H$\alpha$ (M.\ Ehle, private communication).

In the following we discuss the spectral signature in the soft X-rays of 
various dynamically and thermally self-consistent galactic wind flows.

\subsection{Global winds from spiral galaxies (slow winds)}
As far as the general properties of global winds are concerned we refer to 
the parameter studies for adiabatic flows performed by Breit\-schwerdt et 
al.\ (1991). Here we concentrate on the effect of cooling and the spectral
characteristics of such outflows. 

First of all we note that radiative cooling is relevant for global winds.
An estimate of the cooling time gives $\tau_{\rm c} \sim 3 k_B T/(n \Lambda) 
\approx 3.9 \times 10^7$ yrs, using the isochoric cooling function 
$\Lambda = 1.7 \times 10^{-22} \, {\rm erg} \, {\rm cm}^3 \,{\rm s}^{-1}$ for 
an initial gas temperature of $T=10^6$ K (Schmutzler \& Tscharnuter 1993)
and a gas density of $n = 10^{-3} \, {\rm cm}^{-3}$. 
This value has to be compared to the flow time scale $\tau_{\rm flow} = 
\int_{z_0}^{z_c} dz/u(z)$, 
where the integration runs from the inner 
boundary near the disk (reference level $z_0 = 1 \, {\rm kpc}$) to the 
critical point $z_c$, because at distances $z > z_c$, the stationary flow is causally 
disconnected from the boundary conditions. For values labeled as reference
model in Breitschwerdt et al.\ (1991) and here denoted as model M1
($\pgn = 2.8 \times 10^{-13} 
\dyn \, {\rm cm}^{-2}$, $\pcrn = 1.0 \times 10^{-13} \dyn \,{\rm cm}^{-2}$,
$\rho_0 = 1.67 \times 10^{-27} {\rm g} \,{\rm cm}^{-3}$,
$B(z=z_0) = B_0 = 1.0\/ \, \mu{\rm G }$ and
$\alpha_0:= |\langle\delta\vec{B}_0 \rangle|/|\vec{B}_0| = 0.1$, corresponding 
to a negligible wave pressure $\pwn = 4.0 \times 10^{-16} \dyn \,
{\rm cm}^{-2}$), we find from the numerical calculations $\tau_{\rm flow} = 
8 \times 10^8 \, {\rm yrs}$. We note that this time scale is an order of 
magnitude larger than the average superbubble lifetime. Therefore global 
galactic winds should be energetically sustained by a large ensemble of disk 
superbubbles rather than by individual star forming regions, as it is 
the case for local winds. 

\begin{figure}[htbp]
      \psfig{file=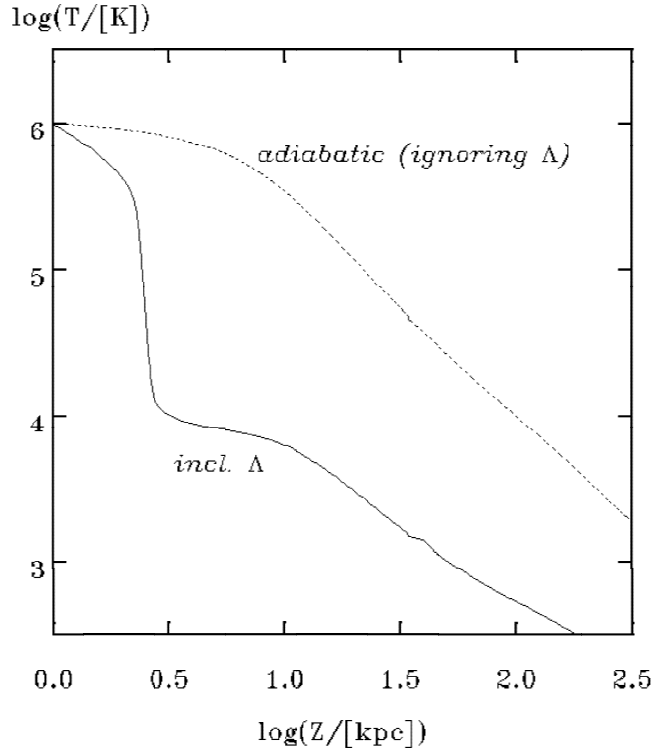,width=\hsize,clip=}
      \caption[]{Temperature profile for a global galactic wind, using
               boundary conditions, appropriate to the reference model M1
               (see text). The dashed line corresponds to the adiabatic
               model, whereas the solid line is obtained by including 
               the isochoric cooling function $\Lambda$. 
              }
         \label{Fig-rm-T1}
   \end{figure}

In the following we discuss the major results of our numerical simulations. 
The effect of adiabatic and radiative cooling 
can be studied most directly by comparing the 
respective temperature profiles (see Fig.~\ref{Fig-rm-T1}).
%
%
%
In the adiabatic model, cooling by $P\, dV$-work becomes noticeable only 
at distances $z \approx 7 \, \kpc$ from the disk. Up to $z \approx 30\, \kpc$ 
the temperature is roughly a power law $T \propto z^{\alpha}$ with index 
$\alpha = -1.59$ and at even larger distances increasing to $\alpha = -1.38$. 
The latter value is close to the expected value of 
$\alpha = -1.33$, because far out in the halo the wind velocity is close to its
asymptotic value, and according to Eq.~(\ref{dyn-eq-13a}), $\rho \propto z^{-2}$.
In an adiabatically expanding flow, $T \propto \rho^{\gamma - 1} = z^{-4/3}$, for 
$\gamma=5/3$. 

Including radiative cooling, we find that already at $z = 3\, \kpc$, there is 
a sharp decrease in temperature from $3 \times 10^5 \, {\rm K}$ to almost 
$10^4 \, {\rm K}$ within one kiloparsec. The reason for this behaviour lies in
the low initial flow velocity of around 10~km/s, which leads to a flow time of
about $1.8 \times 10^8 \, {\rm yrs}$. The corresponding isochoric cooling time 
scale for a gas at $10^6 \, {\rm K}$ is $\tau_c \sim 3 \, k_{\rm B} \, 
T/(n \, \Lambda) \approx 3.9 \times 10^7\, {\rm yrs}$ for $n = 10^{-3} \, 
{\rm cm}^{-3}$ and $\Lambda = 1.7 \times 10^{-22}\,{\rm erg}\, {\rm cm}^3 
\, {\rm s}^{-1}$.
Since the density has decreased by only 26\% up to $z = 3\, \kpc$,
the isochoric cooling function is not too bad an approximation.
 
\begin{figure}[htbp]
       \psfig{file=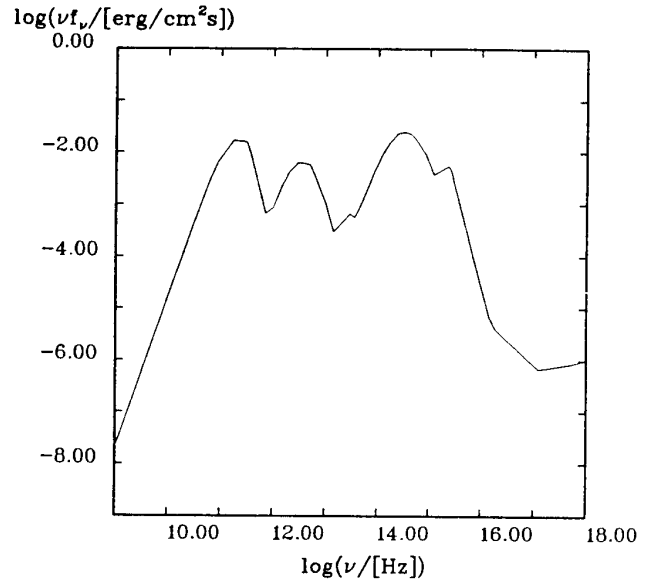,width=\hsize,clip=}
      \caption[]{Spectral energy flux of the diffuse radiation field in the Galactic 
               halo. The curve has been obtained by using data from Black (1987) 
               and Dorfi (1992).
              }
         \label{Fig-rm-S}
   \end{figure}

It is worth pointing out, that due to recombination, the degree of ionization, 
defined as
\beq 
x = 1 -  {\sum_{Z}^{} n_{Z,0} \over \sum_{Z,z}^{} n_{Z,z}} \,,
\label{id-1}
\eeq
where $n_{Z,z}$ denotes a $z$ times ionized atom (not to be confused with the 
distance variable $z$) with nuclear charge $Z$, drops from 1.0 to 0.9 at a 
distance of 3 kpc already. The corresponding amount of neutrals is then 
sufficiently large 
to damp away the self-excited waves by ion-neutral damping (Kulsrud \& Pearce 
1969). Near the sonic point of the flow, we have $x \approx 0.5$. This would 
lead to a redistribution of energy from the CRs to the thermal plasma with 
subsequent losses due to radiative cooling. 
However, it is has been argued (Breitschwerdt et al.\ 1991), that a diffuse
radiation field from stars in the Galactic disk and also from extragalactic 
sources (AGNs, quasars, etc.) could provide the necessary photoionization 
in order to keep the halo fully ionized. We have used the data collected by 
Black (1987) and Dorfi (1992) and have closed the UV gap by spline interpolation, 
resulting from absorption in the Galactic disk, by a power law. The form of the 
spectrum is shown in Fig.~\ref{Fig-rm-S}.
%
%
%
The diffuse photon field acts also as a heat source $\Gamma$, which has to 
be incorporated along with the cooling function $\Lambda$ in a self-consistent 
fashion, as it is shown below. 
It is included in the following calculations under the assumption that its spectral 
energy flux is independent of the vertical distance from the disk. 
This seems reasonable because the disk acts as an extended radiating surface, and 
therefore the energy flux remains constant for distances comparable to the diameter 
of the galaxy, and the extragalactic contribution is constant everywhere.
It turns out, that $x$ is always sufficiently close to unity in 
order to suppress linear wave damping. However the energy input by photoionization
cannot compensate for the heavy radiative losses between 1 and 3 kpc. 
The photon field gains influence at lower densities and is able to compensate 
for radiative but not for adiabatic losses.  
This analysis clearly shows the necessity for performing dynamically and 
thermally self-consistent calculations along the lines 
described in Section~2. 
The variables, which we use for iteration are the 
{\it dynamical} and the {\it thermal} temperature, $T_{\rm d}$ and 
$T_{\rm t}$, respectively, as has been discussed in Sect.\ 2.4.

\begin{figure}[htbp]
      \psfig{file=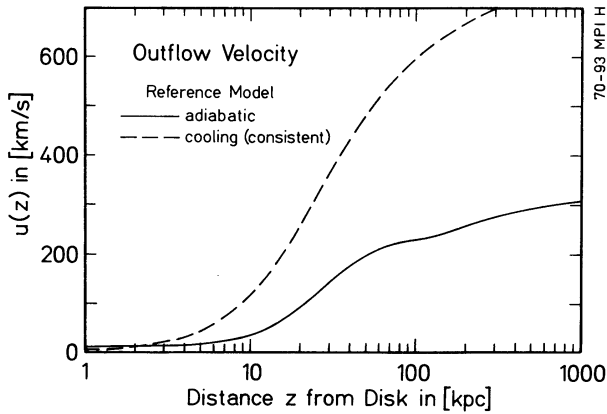,width=\hsize,clip=}
      \caption[]{Outflow velocity of an adiabatic flow (solid line) and a dynamically 
       and thermally self-consistent flow with cooling (dashed line) for a 
	{\it global} galactic wind, using 
               boundary conditions, appropriate to the reference model M1 
               (see text).
              }
         \label{Fig-rm-u}
   \end{figure}

\begin{figure}[htbp]                                                    
      \psfig{file=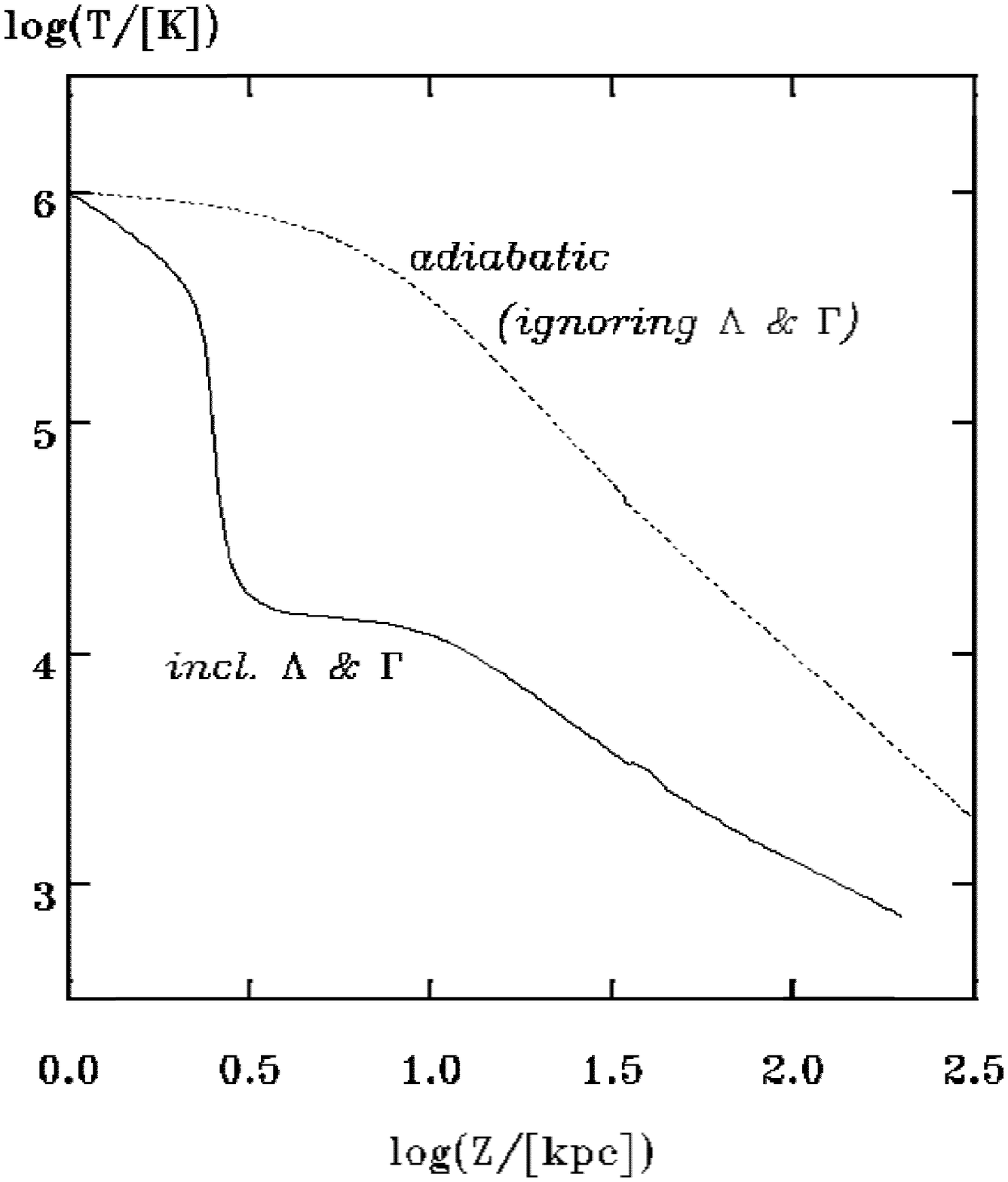,width=\hsize,clip=}                     
      \caption[]{Temperature profile for a global galactic wind, using     
               boundary conditions, appropriate to the reference model M1     
               (see text).                                                 
               The dashed line corresponds to the adiabatic                
               model, whereas the solid line is obtained by including      
               the isochoric cooling function $\Lambda$ and, in comparison 
               to Fig.~\ref{Fig-rm-T1}, also                               
               heating by the external photon field.                       
              }                                                            
         \label{Fig-rm-T2}                                                 
   \end{figure}

\begin{figure}[htbp]
      \psfig{file=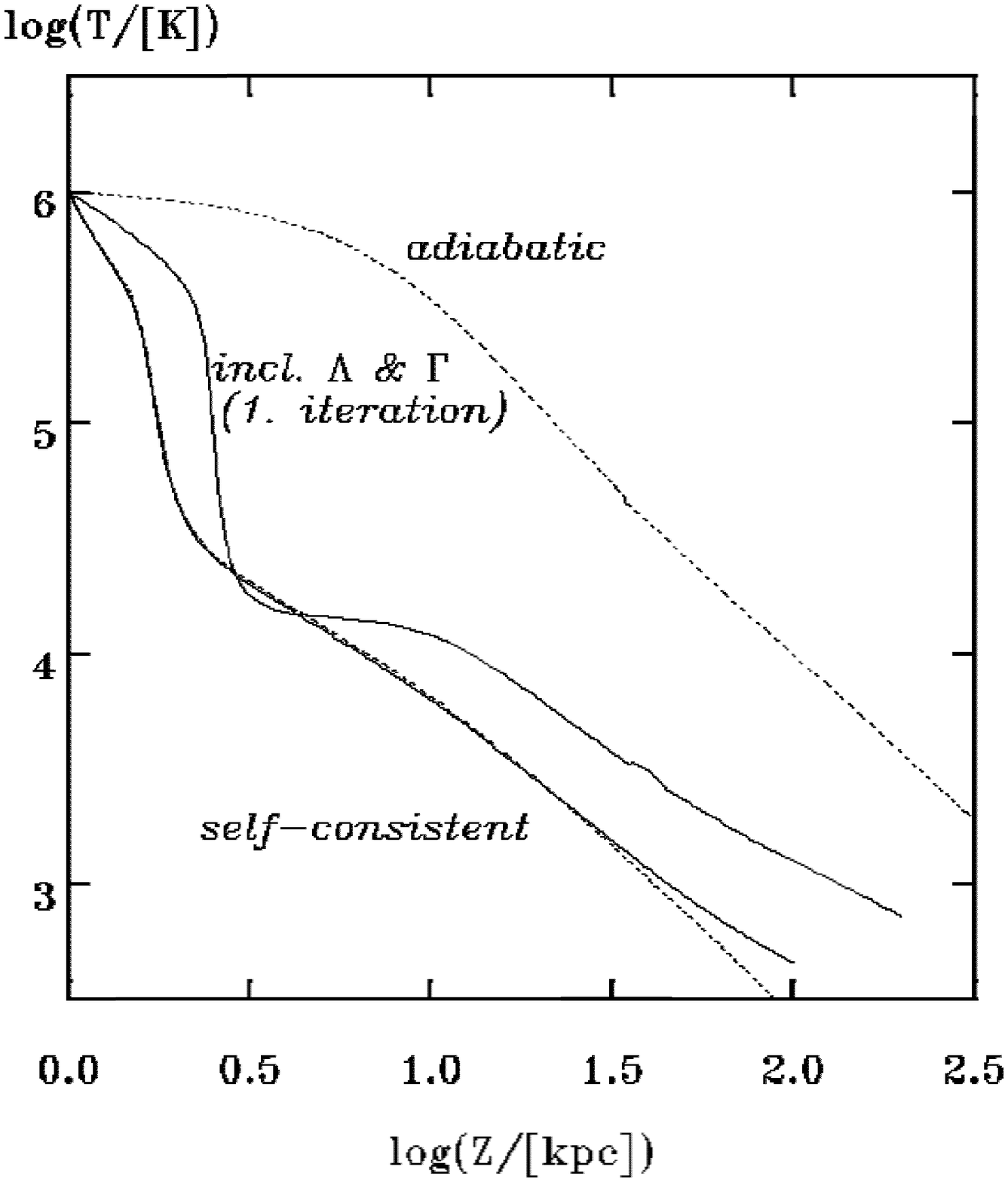,width=\hsize,clip=}
      \caption{Temperature profile for a global galactic wind, using
               boundary conditions, appropriate to the reference model M1
               (see text). The dashed upper line corresponds to the adiabatic 
               model, whereas the dashed lower line shows the self-consistent 
               thermal temperature profile, which deviates from the 
               dynamical temperature profile (lower solid line) by only a
               small amount at distances larger than 30~kpc. The upper solid 
               line shows the thermal temperature profile for an intermediate 
               iteration. 
              }
         \label{Fig-rm-T3}
   \end{figure}

In the following we will present our results. 
The reduced 
initial velocity of the flow ($u_0 = 3.7 \, {\rm km} \, {\rm s}^{-1}$), and 
hence a correspondingly reduced mass loss rate by a factor of 2.5
(see Fig.~\ref{Fig-rm-u}), 
gives rise to a substantial acceleration, and higher velocities at large
distances from the Galactic plane. This fact can be attributed to the very 
large scale height of the CRs, which are unaffected by radiative 
cooling. Therefore their energy density (together with the wave energy 
density) leads to a significant acceleration of the reduced wind mass further
out in the flow. With
increasing flow velocity, adiabatic cooling dominates and the slopes of the
temperature curves (adiabatic versus radiative cooling) in 
Fig.~\ref{Fig-rm-T2} and Fig.~\ref{Fig-rm-T3} are similar. 
%
%
%
%
%
The new temperature profile as a function of distance from the disk is shown 
in Fig.~\ref{Fig-rm-T3}. 
%
%
%
As a result of the low initial velocity, the flow time becomes rather large 
and therefore radiative cooling close to the inner boundary is very 
efficient. From Fig.~\ref{Fig-rm-T3} we infer that already at $z = 1.5 \, 
{\rm kpc}$, line cooling dominates and the temperature drops from the 
initial value of $T_0 = 10^6 \, {\rm K}$ by more than 
one order of magnitude within the next kiloparsec. 
Therefore the resulting \textit{self-consistent} temperature profile is 
much closer to an isochorically than to an adiabatically cooling gas. 
When the temperature is 
as low as only a few $10^4$ K, adiabatic cooling dominates 
line cooling. 

\subsection{Local winds from spiral galaxies (fast winds)} 
\begin{figure}[htbp]
      \psfig{file=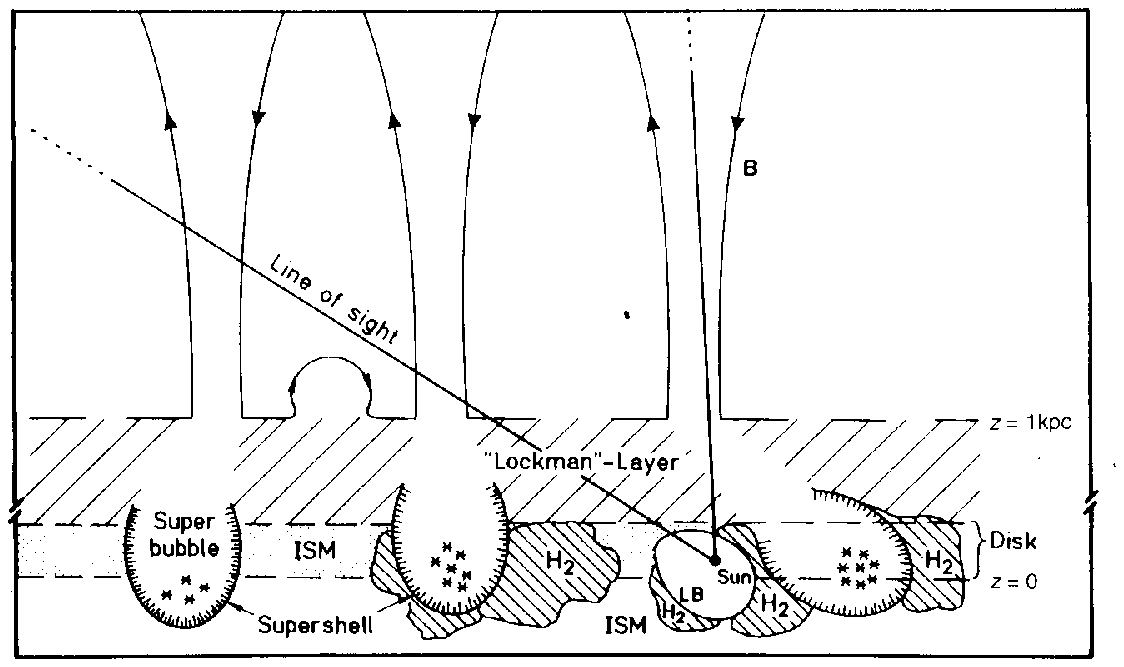,width=\hsize,clip=}
      \caption[]{Schematic edge-on view of the Galaxy (not drawn to scale) 
       taken from Breitschwerdt \& Schmutzler (1994).
                     }
         \label{Fig-sketch1}
   \end{figure}

Observationally, the concept of local winds is supported by large scale 
filaments ($\sim 1 \, {\rm kpc}$) protruding out into the galactic halo, 
as it is the case for a number of inclined spiral galaxies, like 
NGC$\,$891, NGC$\,$5775 and also the Milky Way (Dettmar 1992). The 
observation of vertical dust lanes may be interpreted as the dense walls 
of chimneys sticking out of the Galactic disk (Sofue 1991; Sofue et al., 1991).
The analysis of the H\,{\sc i} distribution in galaxies reveals the existence of 
supershells (Heiles 1979; 1984) as in our Galaxy, or distinct holes in 
H\,{\sc i} like in M31 (Brinks \& Shane 1984). Assuming that H\,{\sc i} and hot gas are
anticorrelated in the ISM on large scales, since the latter displaces the 
former, these findings suggest the existence of superbubbles with typical 
diameters between 100 and 1000 pc and a time integrated energy content of 
$10^{52}$--$10^{53} {\rm erg}$. We have argued earlier, that these 
superbubbles are susceptible to break-out of the disk, most likely where 
the surface density is lowest, and thus feed the halo. It is therefore 
possible, that a flux tube in the direction of the north Galactic pole 
may have also been created by a nearby superbubble. 
In Fig.~\ref{Fig-sketch1} we have sketched a plausible situation for 
our Galaxy, which we assume to be typical 
for other spirals as well, if observed edge-on.  
%
%
%
%
Considering the huge amount of energy stored in superbubbles, which
is mainly in thermal, kinetic and CR energy, it is 
clear, that the boundary conditions for {\it localized} outflows must be 
quite different from the global outflows discussed earlier. 

\begin{figure}[htbp]
      \psfig{file=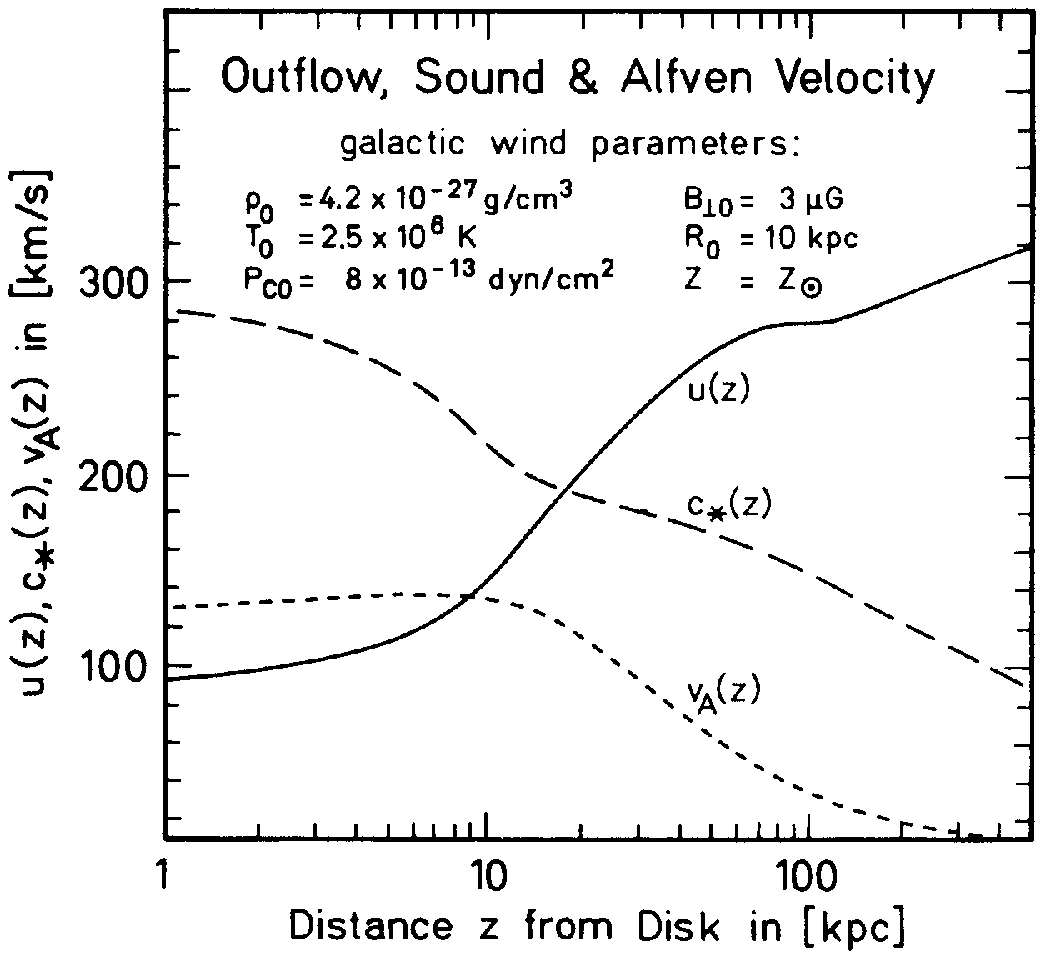,width=\hsize,clip=}
      \caption[]{Outflow velocity $u(z)$, Alfv\'en velocity $v_{\rm A}(z)$ 
        and ``compound sound speed'' $c_*(z)$ (cf.\ Eq.~(\ref{dyn-eq-20})) for a 
	{\it local} galactic wind, with suitable boundary conditions 
	(model M2), (see text).               
              }
         \label{Fig-lw-u}
   \end{figure}

\begin{figure}[htbp]
     \psfig{file=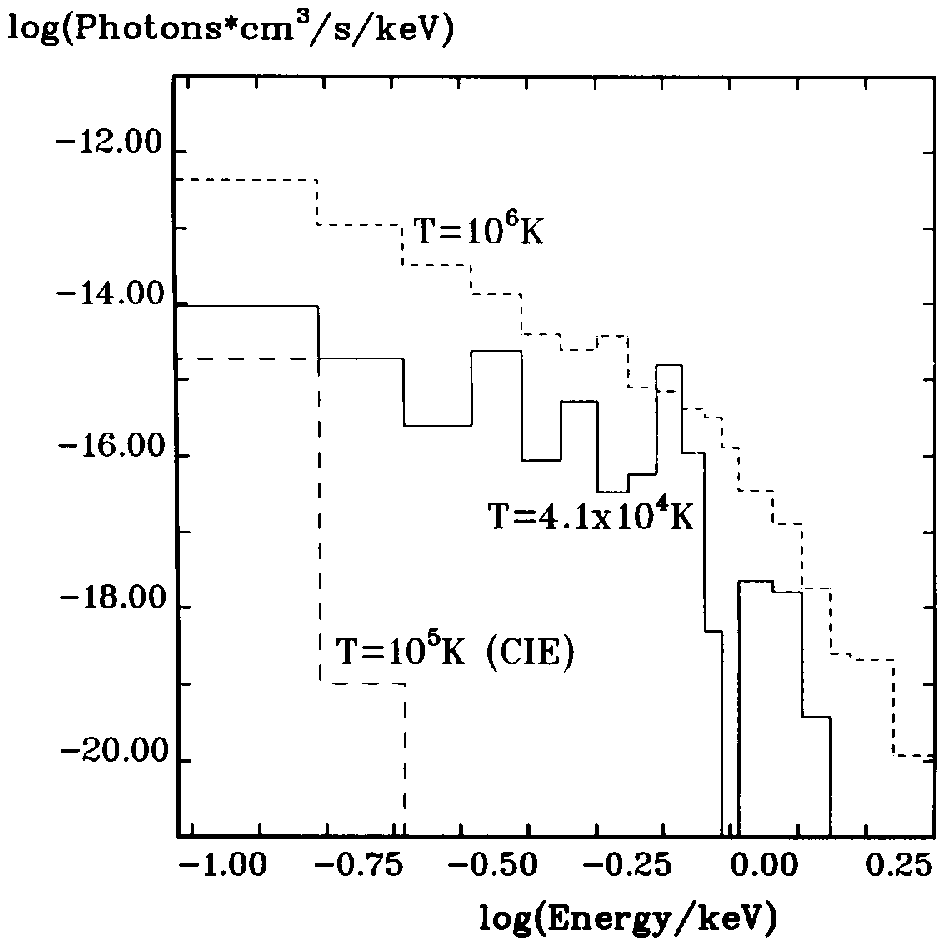,width=\hsize,clip=}
      \caption[]{Photon spectra for local emission from a local galactic wind, 
               normalized to ${n_{\rm e}}^2$ (for boundary conditions 
	       of model M2, see text). 
               The energy range corresponds to the {\sc Rosat} PSPC instrument;
               channel binning is arbitrary. The short dashed line shows a 
               non-equilibrium  spectrum at $T=10^6 \, {\rm K}$ (i.e.\ 
               $|z| \approx 10 \, {\rm kpc}$), the solid line represents 
               a $T=4.1 \times 10^4 \, {\rm K}$ non-equilibrium spectrum 
               ($|z| \approx 65 \, {\rm kpc}$),  
               and the long dashed line represents a $T=10^5 \, {\rm K}$ 
               CIE spectrum for comparison.                   
              }
         \label{Fig-xrb-sp1}
   \end{figure}

\begin{figure}[htbp]
     \psfig{file=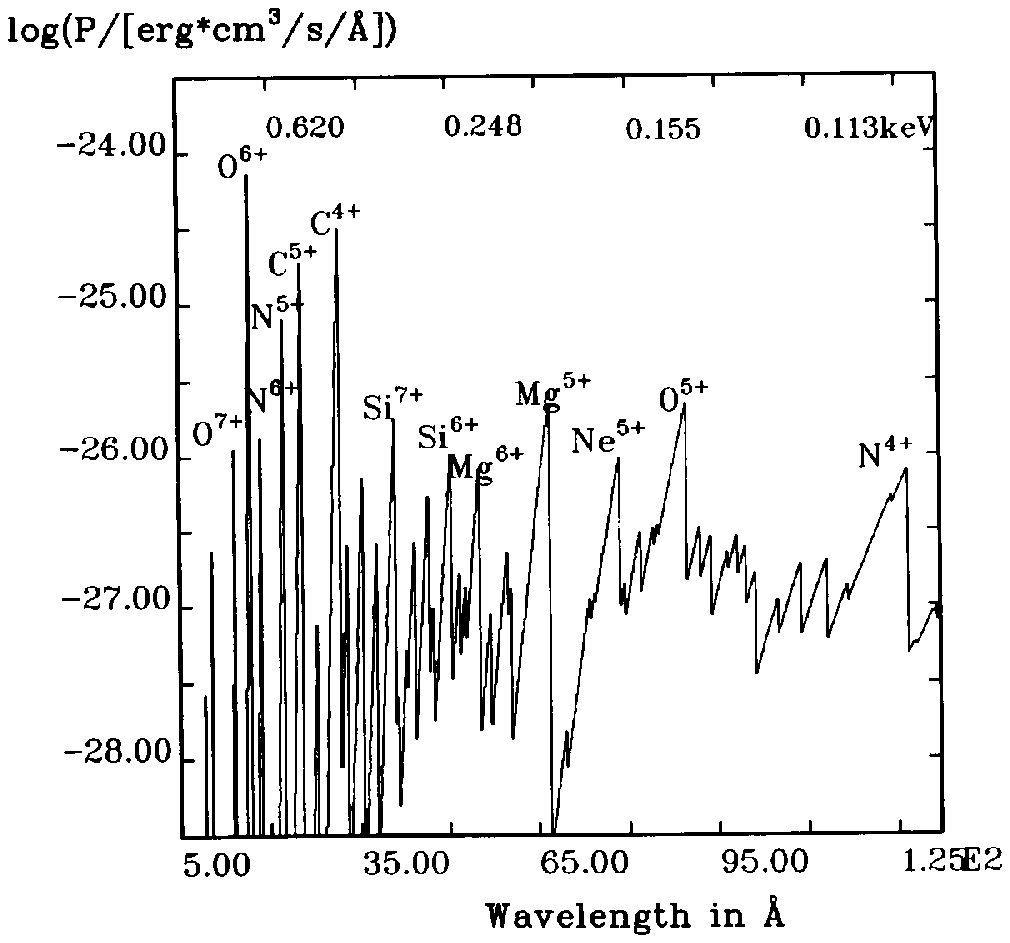,width=\hsize,clip=}
      \caption[]{High resolution photon spectrum for the local emission from 
               a local galactic wind, 
               normalized to ${n_{\rm e}}^2$ (cf.\ Fig.~\ref{Fig-xrb-sp1}; for 
               boundary conditions of model M2, see text).                    
              }
         \label{Fig-xrb-sp2}
   \end{figure}

As a typical 
example, we have chosen the following input values for the dynamical 
model (M2): 
$\rho_0=4.2\times 10^{-27}\, {\rm g} \, {\rm cm}^{-3}$, initial temperature 
$T_0=2.5\times 10^6 \,{\rm K}$, CR pressure $P_{C0}
= 8.0\times 10^{-13}\,{\rm dyn} \, {\rm cm}^{-2}$, regular vertical magnetic 
field component $B_0=3 \mu{\rm G}$ and a  fluctuating magnetic field of 
$\delta B_0 = 0.1 \, B_0$. 
The halo plasma consists of the 10 most abundant elements (cf.\ 
Section 2.3) with solar metallicity. As discussed earlier, we have assumed that 
the gas is in CIE,
because of the high initial temperature due to SN heating. 
However, as we have already pointed out, below $10^6 \, {\rm K}$ this 
assumption is no longer justified.
In Fig.~\ref{Fig-lw-u} the resulting velocity profiles of dynamically and
thermally self-consistent solutions are shown. 
%
%
%
The initial velocity $u_0 = 93 \, {\rm km} \, {\rm s}^{-1}$ of the convergent 
solution, is 
a factor of more than 25 above that for a global wind. The sonic point of  
the flow is located at $z = 17.8 \, {\rm kpc}$, a factor of 2 closer to 
the disk than in the global wind case, which is simply because of the higher 
energy input into the flow and the higher acceleration. The mass 
loss rate is $\dot M = 5.8 \times 10^{-3} \, \msol \, {\rm kpc}^{-2} \, 
{\rm yr}^{-1}$.
The crucial point
is now, that despite a decrease in temperature by about two orders of 
magnitude out to $z \approx 65 \, {\rm kpc}$ due to {\it fast adiabatic} and 
some radiative cooling, many of the initial high ionization stages remain
unchanged. {\it Delayed recombination} of these stages leads to X-ray emission 
predominantly in the so-called M-band (0.5--1.1 keV).
This can be directly seen from Fig.~\ref{Fig-xrb-sp1}, where we have 
shown the intrinsic spectra at different temperatures, corresponding to
different distances in the flow. While the spectrum at $T=10^6 \, {\rm 
K}$ is still similar to the initial spectrum at $T=2.5 \times 10^6 \, 
{\rm K}$ in CIE, the spectrum at $T=4.1 \times 10^4 \, {\rm K}$ does not change 
dramatically and still looks more 
like the $T=10^6 \, {\rm K}$ spectrum than a corresponding CIE spectrum. 
In fact, a CIE plasma at $T=4.1 \times 10^4 \, {\rm K}$ would hardly emit any
X-rays, simply because the kinetic temperature of the electrons for the 
excitation of inner levels would be too low. A spectrum at this temperature 
would be barely visible in this representation. For comparison, we have therefore 
plotted a $T=10^5 \, {\rm K}$ CIE spectrum. The spectral difference of 
equilibrium vs.\ non-equilibrium emission of an optically thin plasma is 
indeed striking. 
%
%
%
The major contributors of the highly ionized species to delayed
recombination can be seen in Fig.~\ref{Fig-xrb-sp2}, which shows a high 
resolution spectrum at $T=4.1 \times 10^4 \, {\rm K}$. 
%
%
%
In this representation the recombination edges and the exponential 
decrease towards shorter wavelenghts become apparent. 

In order to compare these calculations with observations of the soft 
X-ray background (SXRB), we need to integrate the local emission spectra 
along a line of sight weighted by the local density and include 
interstellar absorption.

\section{Comparison and implications for observations}
\subsection{Contribution to the soft X-ray background}
The observational data of the soft X-ray background obtained by the
Wisconsin survey and the {\sc Rosat} All-Sky Survey have been 
extensively reviewed, e.g.\  
by McCammon \& Sanders (1990) and Snowden (1996), respectively. 
The most important features to explain are:
\begin{itemize}
\item[$\bullet$] { 1/4 keV band (C-band):} 
\begin{itemize}
\item an increase in flux with Galactic latitude by a factor of 2--3
\item a distinct anticorrelation with H\,{\sc i}
\item approximate constancy of the Be (77--$111\,{\rm eV}$) to 
B (130--$188\,{\rm eV}$) band ratio with increasing count rate,  
where a column density of $N_{\rm H} = 10^{19} \, {\rm cm}^{-2}$ represents 
unity optical depth for the Be-band
\item more than 50\%  of the flux in certain directions originates beyond 
400 pc as inferred from {\sc Rosat} PSPC shadowing experiments; for a recent 
discussion we refer to Snoweden et al. (1998)  
\end{itemize}
\item[$\bullet$] { 0.5--1.1 keV band (M-band):}
\begin{itemize}
\item the emission is fairly isotropic in latitude and longitude (slight 
enhancement towards the Galactic center), if bright 
individual sources (such as the Cygnus superbubble, Eridanus cavity, 
Loop I etc.) are subtracted. 
\end{itemize}
\end{itemize}
Before interpreting these results, one should be aware that unity optical depth
at 1/4 keV corresponds to an H\,{\sc i} column density of $N_{\rm H}\sim 10^{20} \,
{\rm cm}^{-2}$ and at 0.5 keV to $N_{\rm H}\sim 10^{21} \, {\rm cm}^{-2}$. 
Taking an average interstellar density of $n_{\rm H\,{\sc i}} \sim 1 \, 
{\rm cm}^{-3}$, the mean free path of a C-band photon is 100 pc and of an 
M-band photon is around 1 kpc. 
%
   \begin{figure}[tphb]
     \psfig{file=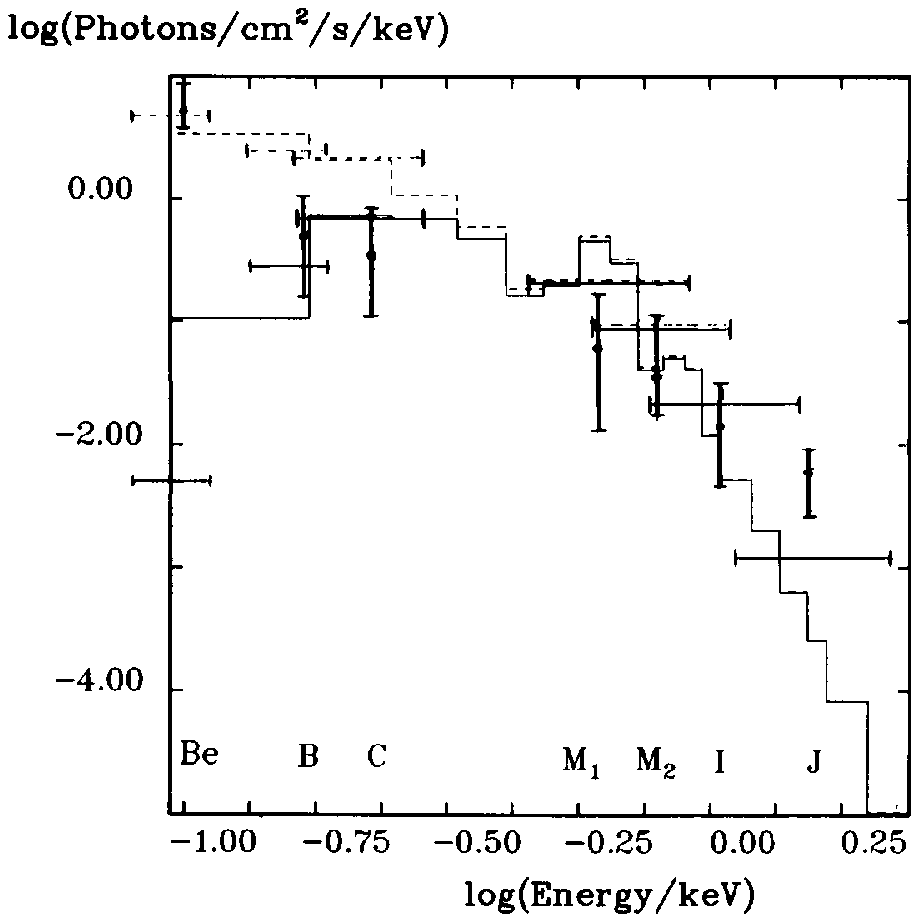,width=\hsize,clip=}
      \caption[]{Photon flux spectra integrated along a flux tube in the Galactic
       wind model (for boundary conditions, see text) with arbitrary binning 
       (cf.\ Breitschwerdt \& Schmutzler, 1994). 
     The solid line represents the flux reduced by an 
     extended H\,{\sc i} layer ($N_{\rm H}  \approx
     1.5 \times 10^{20} \,\rm cm^{-2}$), and the dashed line  
     corresponds to a minimum absorption by the Local Cloud   
     ($N_{\rm H} \approx 3 \times 10^{18}  
     \,\rm cm^{-2}$). The horizontal bars show the flux averaged for the 
     following  energy bands: Be (77--111$\,$eV), B (130--188$\,$eV), C 
     (160--284$\,$eV), M1 (440--930$\,$eV), M2 (600--1100$\,$eV), I 
     (770--1500$\,$eV) and J (1100--2200$\,$eV) bands, 
     including absorption. The vertical bars give the 
     minimum to maximum range of the measured fluxes taken from the Wisconsin
     survey in the same energy bands. 
              }
         \label{Fig-xrb-wisc1}
   \end{figure}
%
%
Therefore for both energy bands, a purely extragalactic origin is ruled out 
regarding the observed emission in the Galactic plane. The pre-{\sc Rosat} 
explanation that the C-band flux is entirely due to the thermal emission
of a local hot ($T\sim 10^6 \, {\rm K}$) gas, displacing the H\,{\sc i} 
(``displace\-ment'' model;  Sanders et al.\ 1977; Snowden et al.\ 1990)
and filling homogeneously a cavity of about 100 pc in radius, has been 
clearly ruled out by the so-called shadowing experiments. 
For example Herbstmeier et al.\ (1995) studied the shadows cast by the
high latitude Complex M of high velocity clouds, which is at a distance of 
at least 1.5 kpc. They conclude, that 1/4 keV band emission also 
originates in a Galactic corona which extends up to several kiloparsecs from 
the plane except for some regions, where it could be less.  
Clearly, a successful 
model {\it must} be capable of explaining ultrasoft X-ray emission from 
beyond the Local Bubble, i.e. from the Galactic halo. 
Whereas the 1/4 keV diffuse X-ray emission is almost entirely thermal in 
origin, there is a contribution of about 60\% (Wang \& McCray 1993) to
the 3/4 keV band from extragalactic point sources. Galactic point sources could 
also provide up to 15\% of the flux (Schmitt \& Snowden 1990). On the
other hand there must be a contribution from the Galactic plane {\it and} 
the halo in order to explain the observed isotropy. 
%
   \begin{figure}[htbp]
     \psfig{file=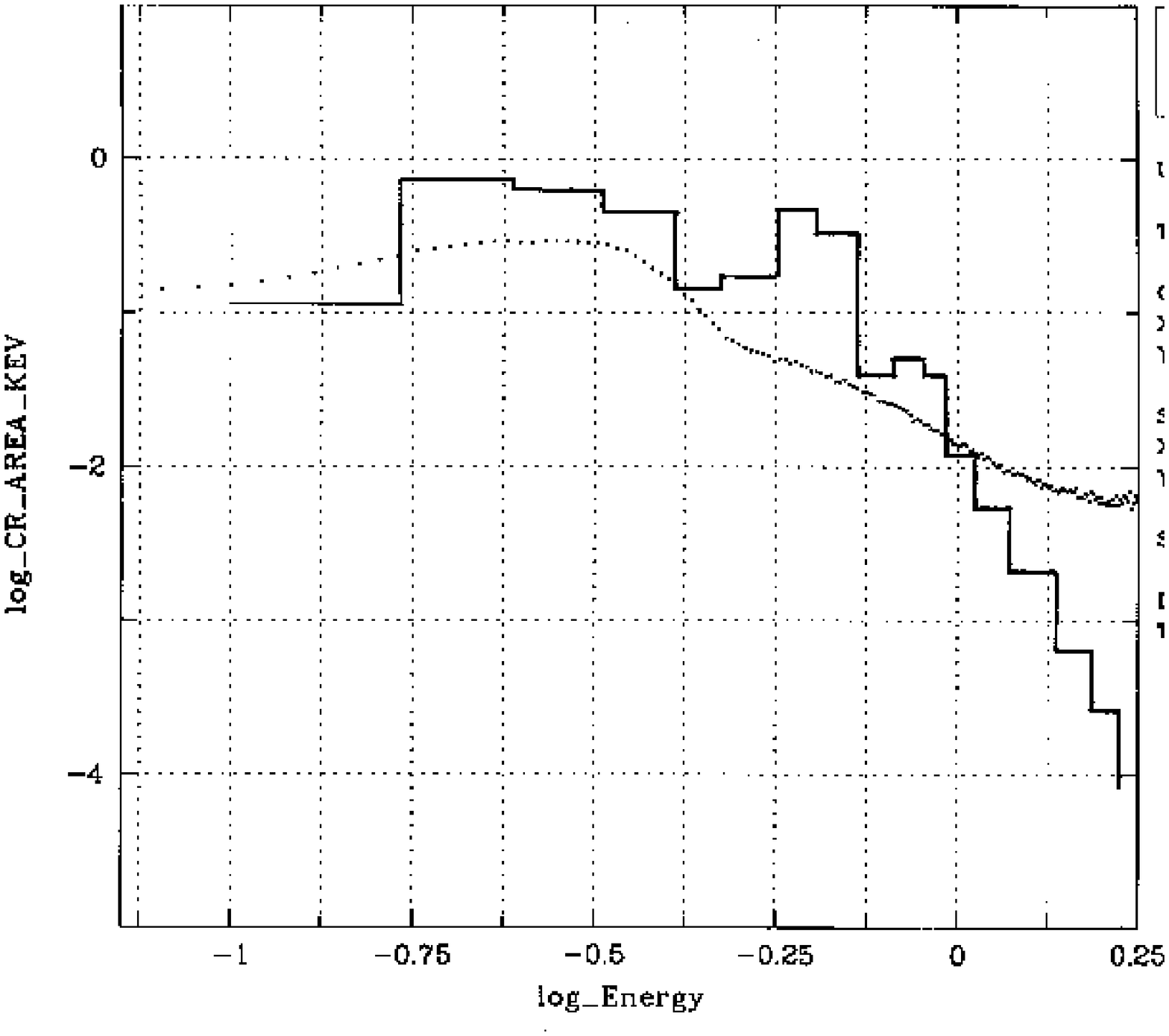,width=\hsize,clip=,angle=90}
      \caption[]{Photon flux spectra integrated along a flux tube in the Galactic
       wind model. Axis labeling and the solid line are the same as in 
       Fig.~\ref{Fig-xrb-wisc1}
       including absorption by the ``Lockman'' layer.   
       The dotted line represents the SXRB as observed in a deep pointed 
       {\sc Rosat} PSPC observation in the direction of the north Galactic pole 
       (courtesy by M.\ Freyberg). 
              }
         \label{Fig-xrb-ros1}
   \end{figure}
%
%
There is no doubt that the presence of the ultrasoft X-ray bands argues 
for a local contribution to the SXRB from the Local Bubble. It is also 
possible that even a small fraction of the emission in the M-band is 
of local origin (Sanders 1993). What is needed in order to substantiate
these statements are shadowing experiments towards high column density 
regions \textit{inside} the Local Bubble, such as the MBM clouds. 
However there are not many suitable targets (like e.g.\ MBM12 and 
MBM16; Magnani et al.\ 1985, 1996; Hobbs et al.\ 1988) that 
fulfill this requirement (Kuntz et al.\ 1997). 
We have shown (Paper I) that fast adiabatic cooling of an expanding
Local Bubble can produce
emission in these energy bands by delayed recombination of highly ionized 
species. Thus the problem of isotropy becomes less severe. 
Future XMM observations will help to clarify the contribution of 
M-band emission within the Local Bubble. The absolute 
M-band value in our models depends of course on the 
initial conditions, in particular on the initial temperature $T_0$.    
The resulting lower temperature in the bubble at the present stage of evolution 
could explain the deficiency of EUV line emission, that was reported from 
observations with the EUVE satellite (Jelinsky et al.\ 1995). 
This deficiency also is one of the results of our model of the Local Bubble
(Paper I) and we have presented a theoretical EUV spectrum in Schmutzler 
\& Breitschwerdt (1996).
For further results and discussion we refer the 
reader to Breitschwerdt (1996) and Breitschwerdt et al.\ (1996) and to a 
forthcoming paper, in which more details concerning the Local Bubble are 
addressed. 

Here we will focus on the contribution to the soft X-ray background by the 
Galactic halo. 
In the following we discuss the spectral properties of diffuse X-ray emission 
in the direction of the north Galactic pole. To that end we have performed 
dynamically and thermally self-consistent calculations of a local outflow 
using the boundary conditions described in the previous section (cf.\ also 
Fig.~\ref{Fig-lw-u}) and subsequently integrated the local emission spectra 
along the line of sight. Note that in a Raymond \& Smith emission model,
the halo spectrum would be described by a single temperature, whereas we 
have a ``multi-temperature'' plasma in which the emission is in addition out of 
equilibrium at each kinetic temperature. 

Let us briefly examine the power radiated away in X-rays. There has been 
criticism (Cox 1998) that the non-equilibrium emission from the Galactic 
halo as a result of an outflow (fountain or wind) would be insignificant in 
comparison to the bulk of X-rays coming from the disk-halo interface, which 
is in CIE. The line of argument is as follows. 
Recombination continuum is mainly due to heavy ions 
which have an abundance of $10^{-3}$ relative to hydrogen. 
Each recombination gives an energy loss of $10^{-3} \times 1 \, 
{\rm keV/recombination} = 2 \times 10^{-12} \, {\rm erg/H-atom}$. 
Taking a mass loss rate of $0.4 \msolyr$ averaged over the whole disk, 
gives $0.4 \msolyr \times 10^{57} \, {\rm atoms}/\msol = 4 \times 10^{56} 
\, {\rm atoms} \, {\rm yr}^{-1}$ corresponding to an energy loss rate 
of $10^{45} \, {\rm erg} \, {\rm yr}^{-1}$.  
Both sides of the Galactic disk contribute an energy flux of 
$1/(2 \pi \times (10 \, {\rm kpc})^2) \, \times 10^{45} {\rm erg} \, 
{\rm yr}^{-1} = 
6 \times 10^{-9} \, {\rm erg} \, {\rm cm}^{-2} \, {\rm s}^{-1}$,  
which is of the order of a percent of the total measured energy flux from 
the SXRB. 

However, there are two major objections against these arguments. 
Firstly, the mass loss rate of $0.4 \msolyr$ is a \textit{global} value, 
averaged over the whole disk. It could in general easily be larger by a
factor of 10 or so for a local outflow with a large overpressure. 
Secondly and more importantly, for a given initial temperature 
$T_0 = 2.5 \times 10^6 \, {\rm K}$ 
we have roughly $250 \, {\rm eV/atom}$ of thermal energy; part of it is used for 
fast adiabatic expansion and another part for line cooling of collisionally 
excited lines. 
The crucial difference between quasi CIE models and the non-equilibrium emission 
model is that due to the expansion                                                                 
of the wind, the temperature is a function of time or distance $z$ from the disk 
(in a steady state). Thus the line cooling which produces a significant 
contribution at lower heights cannot be described by a Raymond \& Smith model,
because line cooling occurs at ``different temperatures'' as a function of 
distance from the plane; 
moreover, the emerging spectrum cannot be modeled by a superposition of a 
series of 
Raymond \& Smith spectra at different temperatures. Even if the wind starts with 
a gas described by CIE, the ionization states differ more and more from 
those of CIE models with increasing distance from the disk. 

Another way to look at it is the following. In the above argument by Cox it is 
assumed that recombination takes place instantaneously, or on a time scale 
comparable to that of the mass loss. However, at least for local winds this is 
definitely not the case. Our calculations show that recombination occurs very slowly 
if compared to the dynamical time scale, which means that upon integration along 
a line of sight we have to consider a time scale of about $8 \times 10^7$ years, 
which corresponds to a vertical distance of the flow of 25 kpc from the disk. 
Therefore we see a larger amount of mass (of ions ``frozen'' into their ionization 
stages) recombining than the mass loss rate would indicate. This happens at the 
expense of a lower recombination rate at the beginning when the flow is set up. 
In a steady state model this requires that the flux tube must have reached a 
certain height above the disk, before delayed recombination becomes a significant 
contributor to the soft X-ray emission.

Note that in the model discussed here, at 
$z=1 \, {\rm kpc}$ the initial velocity $u_0 = 93 \, {\rm km} \, {\rm s}^{-1}$ 
is quite significant. Moreover, the density as a function of $z$ is different 
from a CIE model, corresponding also to a relation for $T(z)$. 
The situation is different for the Local Bubble: here we observe a 
snapshot at time $t_0$, i.e. at present, and do not integrate through 
the ``historical path'' of the plasma. Therefore, in this case the 
observed spectrum is entirely dominated by recombination continuum 
(Paper I).

In the calculations here, we have assumed 
that the flux tube lies within the $2^\circ$ opening angle of the {\sc Rosat} PSPC. 
Attenuation of the X-ray flux by the Local Cloud ($N_{\rm H} \approx 3 \times
10^{18} \, {\rm cm}^{-2}$) and/or the extended H\,{\sc i} layer (``Lockman'' layer; 
$N_{\rm H} \approx 1.5 \times 10^{20} \, {\rm cm}^{-2}$) has been taken into   
account (Fig.~\ref{Fig-xrb-wisc1}). 
Our results show, that in the lower part of the halo, the spectrum is 
dominated by line emission. Once the temperature has fallen below $10^6 \, 
{\rm K}$, delayed recombination continuum becomes increasingly important,
in particular in the M-band. We have found that volume elements up to 
a vertical distance of about 25 kpc contribute to the X-ray flux, as can be 
seen from the approximate constancy of the emissivity with decreasing 
temperature in 
Fig.~\ref{Fig-xrb-sp1} in this energy range. Further, our numerical simulations 
show that the density decreases as $\rho \propto z^{-0.5}$ for $|z| \leq 25 \, 
{\rm kpc}$. Let $\Delta z$ be the increment along the axis of the emission 
cone, which has the volume $\Delta V \propto z^2 \Delta z$. The observable 
flux $F$ from the emitting volume is then given by $F \propto \rho^2 
\times V/z^2 \approx 1/z$, for $\Delta z = \rm const$. 

The calculated overall spectral distribution shows a remarkable quantitative
agreement with the Wisconsin survey in the I-band, the M-band, the C-band 
and even the B-band. However it is clear, that most of the observed flux in 
the ultrasoft bands must be of local origin. We note that the model presented
here is capable of reproducing the total observed flux in the various energy 
bands. This is of course unrealistic, because as we have already mentioned, 
there will be contributions from other sources. However, as long as it is not 
clear, what the fraction of emission by these sources is, we do not consider 
it appropriate to present a complex and detailed multicomponent model. 
But we note that our model is conservative, because there is no problem
in explaining a reduced flux by simply relaxing the boundary conditions. 
In addition, the form of the spectrum of the Galactic wind and the Local 
Bubble is not vastly different, so that we need more detailed spectral 
information in order to disentangle both contributions. Deep pointed 
observations by XMM might just be able to provide such data.  

In comparison to earlier instruments, 
there has been considerable improvement in sensitivity and angular resolution 
by the {\sc Rosat} PSPC. In Fig.~\ref{Fig-xrb-ros1} a PSPC spectrum of the 
SXRB in direction of the north Galactic pole is presented. It is taken from 
a deep pointed observation from the {\sc Rosat} archive with all known point 
sources (stars, galaxies, clusters of
galaxies etc.) being subtracted. The photon statistics are excellent, and  
the agreement with our model calculation discussed previously is again 
fairly good, with some deviation below 0.2 keV and above 1 keV.
Below 0.2 keV the PSPC is not ground calibrated, so that data in this 
energy range are inherently uncertain.
In addition the {\sc Rosat} data 
still contain contributions from the foreground, i.e. the Local Bubble, 
and unresolved extragalactic background sources, which have not been removed.
The latter is responsible for most of the flux above 1 keV, hence the higher 
observed flux at this energy. It should be noted however, that our calculations
show a clear M-band excess. This discrepancy may be due to an initial temperature 
in the calculations that is somewhat too large. On the other hand, unless 
higher spectral resolution observations become available, a slight excess cannot 
be ruled out observationally. 

Since the 1/4 keV band emission is mostly thermal in origin and a significant 
fraction in the halo is produced by expanding SNRs, superbubbles 
and Galactic wind, we expect a good deal of variation over the sky. This is 
confirmed by the recent analysis of the {\sc Rosat} All-Sky-Survey data 
(Snowden 1996).  
We emphasize that our model is consistent with the results of the shadowing 
experiments, giving a natural explanation for the 1/4 keV band emission 
in the Galactic halo. In particular, fluctuations may be due to the line of 
sight cutting through different regions of neighbouring flux tubes. 
We have estimated (Paper I) that the number of
flux tubes needed to explain the patchiness in the Wisconsin data is consistent 
with the observed area filling factor of superbubbles (Heiles 1990). 

Direct observational evidence of gas flowing from the disk into the halo 
is still scanty. But X-ray observations tell us that hot gas is the dominant 
ISM component in the halo, at least by volume, and sources in the disk are 
available at large numbers with more than sufficient total energy. 
It is therefore tempting to assume that the flux tube in direction of the 
north Galactic pole in our model calculations is fuelled by gas from 
the neighbouring Loop~I superbubble, 
which has an observed diameter of $\sim 320$ pc and is large enough to 
stick out into the halo. But also the giant radio Loops II and III, which have 
a similar apparent size to Loop~I and hence must be close by, may have been 
good candidates for supplying hot gas in the past (since no observable X-ray 
enhancements are associated with them at present). Clearly, more 
investigations are necessary before we can draw any firm conclusions. 
 
\subsection{Temperature determination of interstellar plasmas}
%
   \begin{figure}[htbp]
     \psfig{file=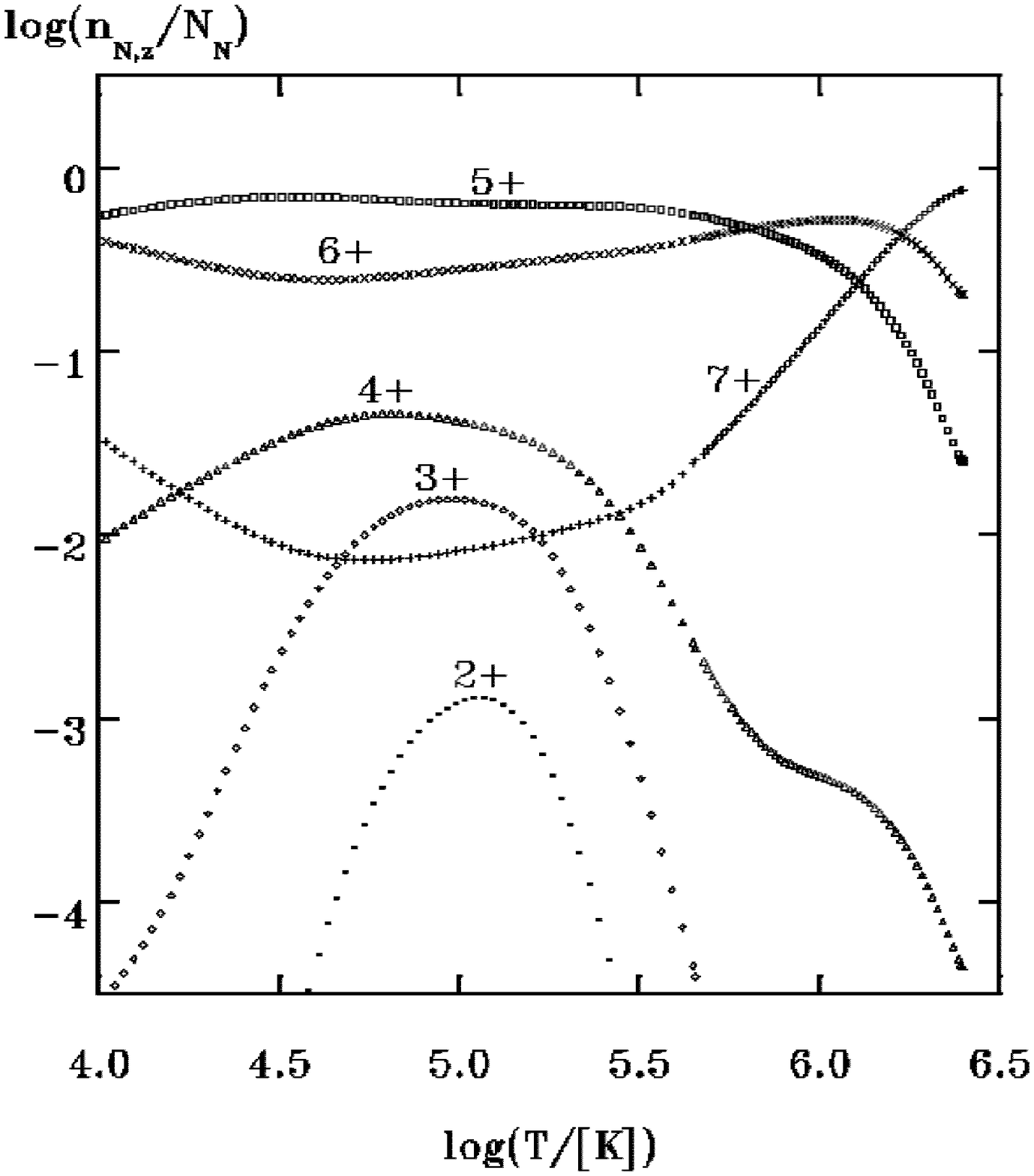,width=\hsize,clip=}
      \caption[]{Ionization state of nitrogen for a plasma 
      not in ionization equilibrium, due to the adiabatic 
      expansion in a galactic wind (for model M2 parameters see text).
      }
         \label{Fig-m2_ni}
   \end{figure}
%
%
%
   \begin{figure}[htbp]
     \psfig{file=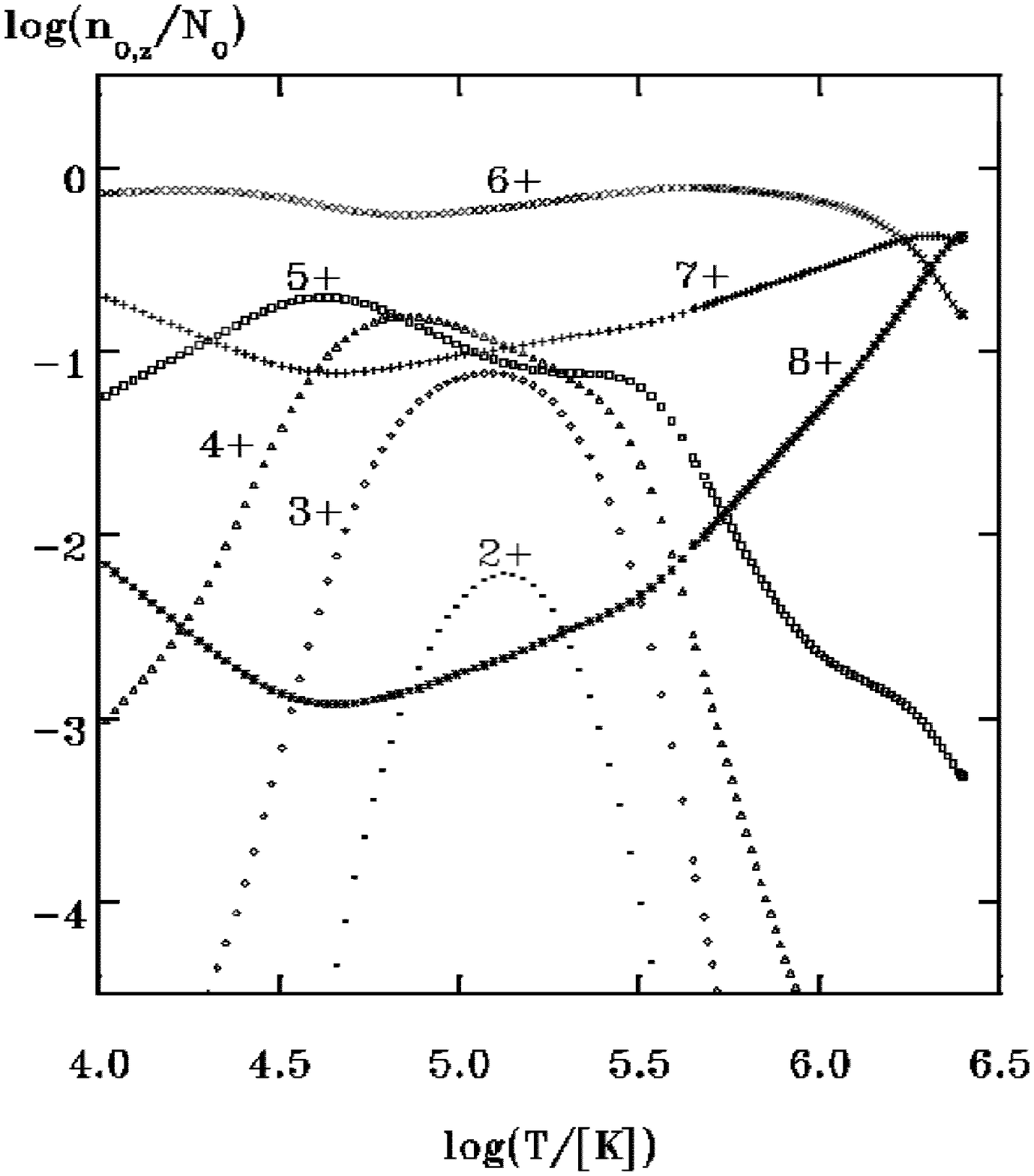,width=\hsize,clip=}
      \caption[]{Ionization state of oxygen for a plasma 
      not in ionization equilibrium, due to the adiabatic 
      expansion in a galactic wind (for model M2 parameters see text).
      }
         \label{Fig-m2_ox}
   \end{figure}
%
%
Highly ionized species such as C\,{\sc iv}, Si\,{\sc iv}, N\,{\sc v} and 
O\,{\sc vi} have been observed in absorption towards 
background objects with a known spectrum both in the Galactic disk 
with the {\sc Copernicus} (e.g.\ Jenkins \& Meloy 1974) and in the halo
with the IUE satellite (Savage \& deBoer 1981; Savage \& Massa 1987;
Danly et al.\ 1992). In order to determine the temperature of the plasma
responsible for the absorption, a convenient method has been to measure 
the ratio of column densities of different lines, e.g.\
$N({\rm N}\,{\sc v})/N({\rm O}\,{\sc vi})$ (York 1974). 
Since there is only a narrow range in temperature, where these two 
absorption lines intersect, the temperature of the plasma is fairly well 
determined. However, this is only true if the plasma is in CIE, 
a fact that has already been noted by York (1974). He mentioned that 
rapid cooling could lead to a lower temperature than would be inferred 
from the line ratios. 
   \begin{figure}[htbp]
     \psfig{file=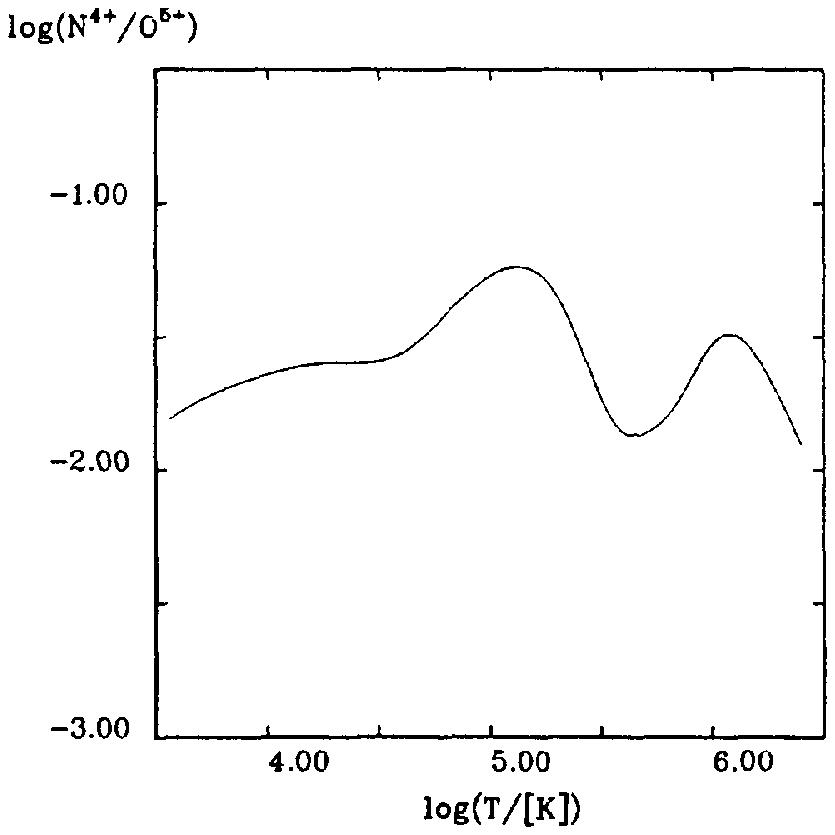,width=\hsize,clip=}
      \caption[]{N\,{\sc v}/O\,{\sc vi} line ratio for a plasma not in 
      ionization equilibrium 
      as a function of temperature in a fast adiabatic galactic wind 
      (for model M2 parameters see text).
      }
         \label{Fig-m2_nv_ovi}
   \end{figure}
%
%
The situation is even more complicated. As a consequence of non-equilibrium 
cooling, the presence of individual highly ionized species depends again 
on the initial conditions and the thermal history of the plasma. 
Moreover, photoionization by an external photon field will also alter the 
ionization stages. To illustrate the effect of delayed recombination, 
we show in Fig.~\ref{Fig-m2_ni} and \ref{Fig-m2_ox} 
the ionization state of nitrogen and oxygen, respectively.
They result from our calculations of a fast adiabatic wind 
(model M2). The presence of highly ionized species at temperatures below 
$T=10^5\,{\rm K}$ is a consequence of recombination delay in a fast cooling plasma. 
The low ionization states at these temperatures are depopulated by 
the external photon field, whose influence grows with the density decrease of the 
outflowing wind. 

It is most instructive to see how the line ratio of 
$\log({\rm N\,{\sc v}}/{\rm O\,{\sc vi}})$ varies with temperature. The most 
interesting feature is that there is no narrow region in which this ratio 
has a peak, but instead there is a rather broad plateau where the line ratio 
stays more or less at the same level. In 
Fig.~\ref{Fig-m2_nv_ovi} we show this line ratio again for model M2.
Therefore caution must be applied if we were to infer temperature 
of interstellar plasmas from line ratios of just 2 species. 
Information on other species is therefore needed, thus constraining 
non-equilibrium ionization models.

\section{Discussion and conclusions}
Most information of the diffuse hot ISM is obtained in the form of
electromagnetic radiation from optically thin plasmas. Under ideal 
conditions of thermal and ionization equilibrium, we can deduce 
the basic parameters of the plasma such as temperature, density 
and pressure \textit{directly} from spectral fitting. However, the 
physical association of hot gas with violent events, such as the 
expansion of individual or multiple SNRs (including their merging)
in the disk and the blow-out of superbubbles into the halo, introduces 
adiabatic cooling which drives the plasma very rapidly out of ionization 
equilibrium. 
As we have shown here, there is a need for a \textit{self-consistent} 
treatment of the 
\textit{dynamical} and \textit{thermal evolution} of the plasma. 
There is no hope, in general, that a convenient CIE spectral fitting procedure 
gives any reasonable results, unless we have reliable information that dynamics 
is unimportant in a particular situation. 

The basic physics behind the dynamical and thermal coupling is that the 
plasma preserves a \textit{memory} of its initial conditions. Moreover, 
the emitted spectrum at any 
time during evolution does depend on the whole thermodynamic path of 
the cooling process. It has been shown (Cox \& Anderson 1982) that the
opposite process of \textit{ionization delay} occurs when an 
adiabatic blast wave runs into an ambient medium which is at lower 
temperature than the hot interior. Once the gas has reached a temperature
in excess of $10^6 \, {\rm K}$, and adiabatic cooling begins to dominate,
then the spectra of SNRs are also modified by recombination delay. We 
have modeled the Local Bubble (Paper I) as a drastic
example, in which a superbubble breaks out of a molecular cloud and the expanding 
shock wave runs down a density gradient. 

On a larger scale, starburst galaxies and AGN winds may be additional candidates 
for a fast adiabatic outflow. As long as there are atoms that can be ionized,
even at high temperatures of the order of $10^7$--$10^8 \,{\rm K}$, delayed 
recombination might play a r\^ole, although the spectrum is largely 
determined by thermal bremsstrahlung. For example the absence of 
the 6.7 keV Fe-K$\alpha$ line -- which should be definitely present at these 
temperatures in a CIE plasma -- in the ASCA spectra of M$\,$82 
(Moran \& Lehnert, 1997) and NGC253 (Ptak et al., 1997), 
can be explained by a fast adiabatically expanding wind.  
We have performed numerical simulations, which will be discussed in detail 
elsewhere, in which the decreasing kinetic energy of the electrons falls rapidly 
below the threshold of collisional excitation of this Fe-line. 
 
In summary however, it is fair to admit that cooling, and also the inverse process 
of heating, of interstellar plasmas are very complicated processes, which are still 
not fully understood and further detailed investigations are necessary.

\acknowledgements
We thank P.\ Biermann, D.\ Cox, C.\ McKee, W.\ Tscharnuter and H.\ V\"olk for
useful discussions. Special thanks to M.\ Freyberg 
who provided unpublished {\sc Rosat} data of the soft X-ray background. 
DB acknowledges support from the 
{\it Deutsche Forschungsgemeinschaft (DFG)} by a Heisenberg fellowship.
The {\sc Rosat} project is supported by the German Bundesministerium f\"ur 
Bildung, Wissenschaft, Forschung und Technologie (BMBF/DLR) and the 
Max-Planck-Gesellschaft (MPG). 
TS did the final part of his work at the Institute for Theoretical
Astrophysics (ITA), University of Heidelberg, supported by a fellowship of the 
DFG (Schm 1025/1 - 1). TS thanks the director of the ITA, W.\ Tscharnuter,
for his warm hospitality and for providing excellent equipment.

\end{document}